\documentclass[twocolumn]{aastex62}
\usepackage{natbib}

\usepackage[hang,raggedright,tight]{subfigure}
\usepackage{longtable}
\usepackage[figuresright]{rotating}
\usepackage{float}
\usepackage[toc,page]{appendix}

\def\gsim{\;\rlap{\lower 2.5pt
 \hbox{$\sim$}}\raise 1.5pt\hbox{$>$}\;}
\def\lsim{\;\rlap{\lower 2.5pt
   \hbox{$\sim$}}\raise 1.5pt\hbox{$<$}\;}

\usepackage{xcolor}

\shorttitle{Constraining Dark Energy With Stacked Concave Lenses}
\shortauthors{Dong et al.}

\begin{document}

\title{Constraining Dark Energy With Stacked Concave Lenses}

\author{Fuyu ~Dong} \affiliation{Department of Astronomy, School of
  Physics and Astronomy, and Shanghai Key Laboratory for Particle
  Physics and Cosmology, Shanghai Jiao Tong University, Shanghai
  200240, China}

\author{Jun ~Zhang*}
\affiliation{Department of Astronomy, School of
  Physics and Astronomy, and Shanghai Key Laboratory for Particle
  Physics and Cosmology, Shanghai Jiao Tong University, Shanghai
  200240, China}

\author{Yu ~Yu}
\affiliation{Department of Astronomy, School of
  Physics and Astronomy, and Shanghai Key Laboratory for Particle
  Physics and Cosmology, Shanghai Jiao Tong University, Shanghai
  200240, China}

\author{Xiaohu ~Yang}
\affiliation{Department of Astronomy, School of
  Physics and Astronomy, and Shanghai Key Laboratory for Particle
  Physics and Cosmology, Shanghai Jiao Tong University, Shanghai
  200240, China}
\affiliation{IFSA Collaborative Innovation Center, and Tsung-Dao Lee
  Institute, Shanghai Jiao Tong University, Shanghai 200240, China}

\author{HeKun~Li}
\affiliation{Department of Astronomy, School of
  Physics and Astronomy, and Shanghai Key Laboratory for Particle
  Physics and Cosmology, Shanghai Jiao Tong University, Shanghai
  200240, China}

\author{Jiaxin~Han}
\affiliation{Department of Astronomy, School of
  Physics and Astronomy, and Shanghai Key Laboratory for Particle
  Physics and Cosmology, Shanghai Jiao Tong University, Shanghai
  200240, China}
\affiliation{Kavli IPMU (WPI), UTIAS, The University of Tokyo, Kashiwa, Chiba 277-8583, Japan}

\author{Wentao~Luo}
\affiliation{Department of Astronomy, School of
  Physics and Astronomy, and Shanghai Key Laboratory for Particle
  Physics and Cosmology, Shanghai Jiao Tong University, Shanghai
  200240, China}

\author{Jiajun~Zhang}
\affiliation{Department of Astronomy, School of
  Physics and Astronomy, and Shanghai Key Laboratory for Particle
  Physics and Cosmology, Shanghai Jiao Tong University, Shanghai
  200240, China}

\author{Liping~Fu} \affiliation{The Shanghai Key Lab for Astrophysics,
  Shanghai Normal University, Shanghai 200234, China}

\email{*betajzhang@sjtu.edu.cn}

\begin{abstract}
  Low density regions are less affected by the nonlinear structure
  formation and baryonic physics. They are ideal places for probing
  the nature of dark energy, a possible explanation for the cosmic
  acceleration.  Unlike void lensing, which requires identifications 
  of individual voids, we study the stacked lensing signals around the 
  low-density-positions (LDP), defined as places that are devoid of foreground
  bright galaxies in projection. The method allows a direct comparison with 
  numerical results by drawing correspondence between the bright galaxies with halos.
  It leads to lensing signals that are significant enough for differentiating several
  dark energy models. In this work, we use the CFHTLenS  catalogue to define LDPs, 
  as well as measuring their background lensing signals. We consider several different 
  definitions of the foreground bright galaxies (redshift range \& magnitude cut).
  Regarding the cosmological model, we run six simulations: the first set of simulations 
  have the same initial conditions,  with $\rm{w_{de}=-1,-0.5,-0.8,-1.2}$;
  the second set of simulations include a slightly different $\Lambda$CDM model and a w(z) model  
  from \cite{2017NatAs...1..627Z}. The lensing results indicate that the models with $\rm{w_{de}=-0.5,-0.8}$ 
  are not favored, and the other four models all achieve comparable agreement with the data.

\end{abstract}


\section{INTRODUCTION}
\label{introduction}

The acceleration of the cosmic expansion remains to be a mystery today
\citep{1998AJ....116.1009R, 1999ApJ...517..565P, 2013PhR...530...87W,
  2014PTEP.2014fB102K, 2014A&amp;A...571A...1P}. It is not yet clear
if it is necessary to go beyond the simplest $\Lambda$CDM model by
introducing a nontrivial equation of state $w(z)$ for dark energy
\citep{1999PhRvD..60h1301H}. Recently, from the baryon acoustic
oscillation measurement of the BOSS data, there has been intriguing
evidence showing a deviation of $w(z)$ from $-1$
\citep{2017NatAs...1..627Z}.  It is desirable to test the nature of dark energy
with alternative cosmological probes. We propose to do so with the
weak lensing effect around low density regions.
 
Low density regions have the advantages of being much less affected by nonlinear evolution
and baryonic physics.  They are likely the ideal places to test dark energy
models with weak lensing. Previous efforts largely focus on
the lensing effect of voids, a typical type of low density region
that is devoid of matter over a significant cosmic volume.

A major challenge of void lensing is about identifying the
voids. Current void-finding algorithms are mostly based on the
distribution of galaxies with spectroscopic
redshifts. \cite{2017MNRAS.465..746S} has summarized these algorithms
into several groups: Watershed Void Finders
\citep{2007MNRAS.380..551P, 2008MNRAS.386.2101N, 2012ApJ...754..109L,
  2015MNRAS.449.3997N}, growth of spherical underdensities
\citep{2002ApJ...566..641H, 2005MNRAS.360..216C, 2005MNRAS.363..977P,
  2006MNRAS.373.1440C, 2011MNRAS.411.2615L}, hybrid methods
\citep{2013MNRAS.434.2167J}, dynamical criteria
\citep{2015MNRAS.448..642E}, and Delaunay Triangulation
\citep{2016MNRAS.459.2670Z}. There are however three main shortcomings
in traditional ways of doing void lensing: 1. void centers cannot be
unambiguously identified due to their intrinsically irregular shapes,
making it difficult to precisely predict or understand the stacked
void lensing signals with a physical model; 2. spectroscopic galaxy
surveys are generally expensive, and suffer from complicated influences
from the selection effects; 3. due to the limited number density of voids
and the scatter of their sizes, the stacked lensing signals do not yet
have a high significance.

More recently, there is a trend to study the lensing effect of low
density regions defined by the projected galaxy distributions
\citep{2015MNRAS.454.3357C, 2017MNRAS.465..746S, 2016MNRAS.455.3367G,
  2017arXiv171005162F, 2017arXiv171005045G, 2017JCAP...02..031B,
  2018arXiv180308717D, 2018MNRAS.481.5189B}.  Comparing to void lensing, these new methods
only need photo-z information, and the stacked lensing signals
generally have much higher significance. For example, \cite{2016MNRAS.455.3367G}
 use a photometrically selected luminous red galaxy sample (redMaGiC) as the 
 foreground galaxies in their paper. By dividing the sky into cells, they assign each
 cell a weighted and smoothed galaxy count.  They do shear measurements around cells 
 with different galaxy counts using DES lensing catalogue. Their follow-up works
 can be found in \cite{2017arXiv171005045G, 2017arXiv171005162F}, in which a
 complete cosmological analysis is presented within  the LCDM models.

Our approach has similarities to their method, but also with differences. We consider these low density 
positions (``LDPs''), which is defined by excluding the foreground bright galaxies from the sky with a critical 
radius in projection. This is a direct way to define the low density regions, 
without defining the galaxy density map.
These positions can be similarly defined in N-body simulations by drawing correspondence
between the foreground bright galaxies and halos/subhalos through, 
e.g., subhalo abundance matching \citep[SHAM, e.g.,][]{2004MNRAS.353..189V}.
These operations are straightforward to realize, and enable us to directly
compare the stacked lensing signals around LDPs with the simulation predictions.
The motivation of this paper is to differentiate several different dark energy models 
 through this type of comparison. 

This paper is organized as follows: In \S\ref{method}, we introduce
the basic theory of weak lensing, and the method for stacking the
lensing signals from low density regions in both observations and
numerical simulations; \S\ref{results} shows our main results for
several different dark energy models; \S\ref{discussion} gives our 
conclusion and discussions about related issues.

\begin{figure}
    \centering
    \subfigure{
    \includegraphics[width=0.98\linewidth, clip]{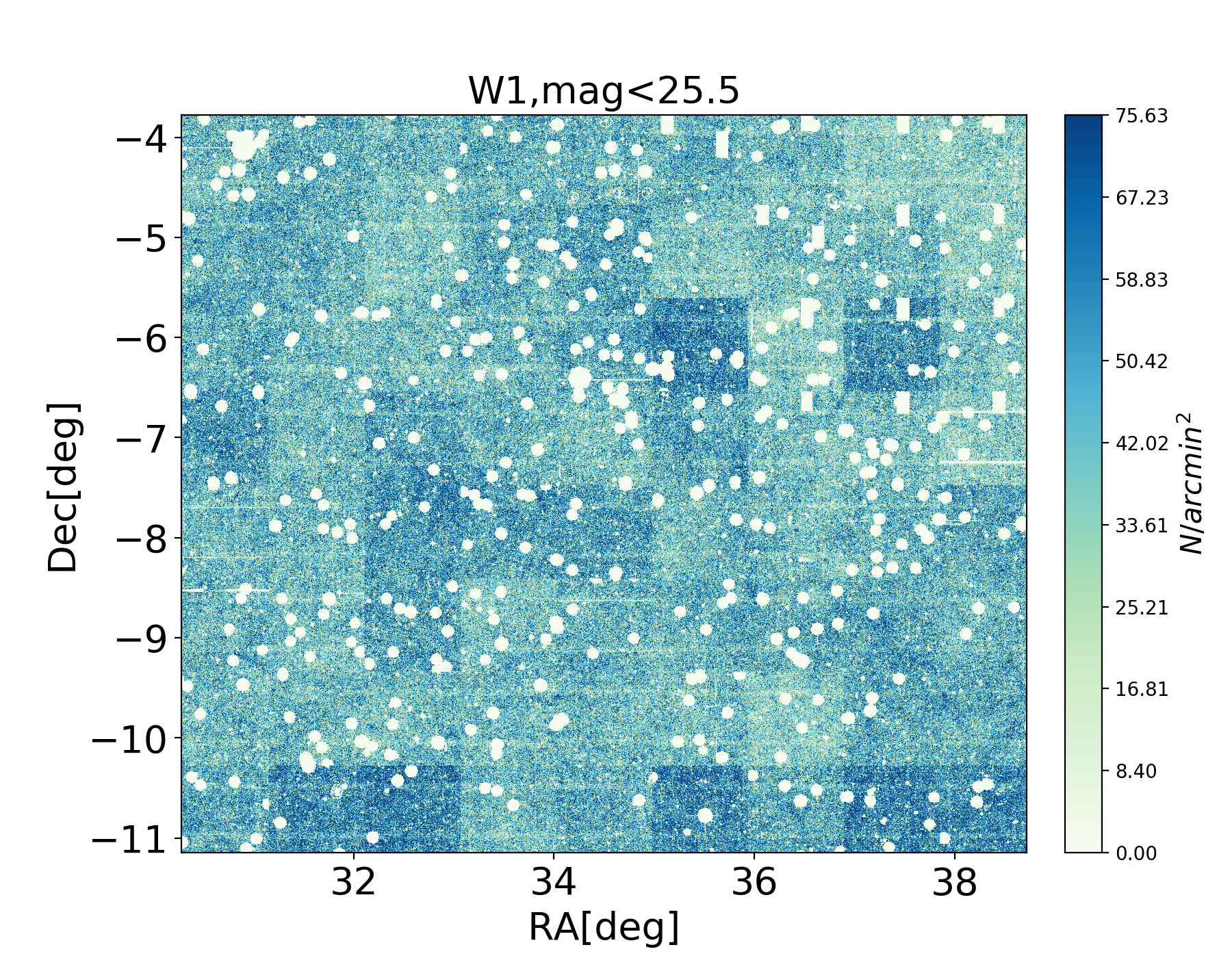}}
  \caption{ Source distribution in W1 from CFHTLenS
    catalogue with apparent magnitude $\mathrm{mag_i<25.5}$.
    The empty areas in this map are masked
  out for bright stars.}
    \label{fig:w1-full}
\end{figure}

\section{METHOD}
\label{method}
\subsection{Overview}

The background tangential
shear is related to the stacked excess surface density of the
foreground (see, e.g., \cite{1998...Peacock}):
\begin{equation}
  \mathrm{ \Delta\Sigma(R)\equiv\Sigma_{cr}(z_l,z_s)
    \langle \gamma_t\rangle(R)=\overline{\Sigma}(<R)-\overline{\Sigma}(R)},
\end{equation}
where R is the distance to the center, and $\rm{\Sigma_{cr}(z_l,z_s)}$ is
the critical surface density in comoving unit, which is defined as:
\begin{equation}
\label{EQ1}
\mathrm{\Sigma_{cr}(z_l,z_s) = \frac{c^2}{4\pi G}\frac{D_A(z_s)}{D_A(z_l)D_A(z_l,z_s)(1+z_l)^2} },
\end{equation}
where $\rm{z_s}$ and $\rm{z_l}$ are the redshifts of the source and the lens 
respectively ($\rm{z_l<z_s}$). $\rm{\overline{\Sigma}(<R)}$ is the
meaning surface density within R, and $\rm{\overline{\Sigma}(R)}$ is
the surface density at R.  $\rm{D_A}$ refers to the angular diameter distance.
By stacking the background shear signals,
Eq.(\ref{EQ1}) allows us to probe the average surface density profile
around the foreground objects (e.g., galaxies, cluster centers, void
centers, etc.) directly, with an enhanced significance and better
circular symmetry.

There are in principle no restrictions on how one defines the
foreground positions as long as they are physically meaningful. For
our purpose of studying the properties of dark energy, we consider
stacking the shear signals around LDPs, which are simply defined
as places that are away from foreground bright galaxies (within a certain narrow
redshift range) by more than a critical distance in projection. 
LDPs defined in such a way generally point to low-density regions. They provide 
abundant foreground positions for shear-stacking,
leading to highly significant lensing signals, as shown in the rest of the paper.


\begin{figure}
    \centering
    \subfigure{
     \includegraphics[width=0.92\linewidth, clip]{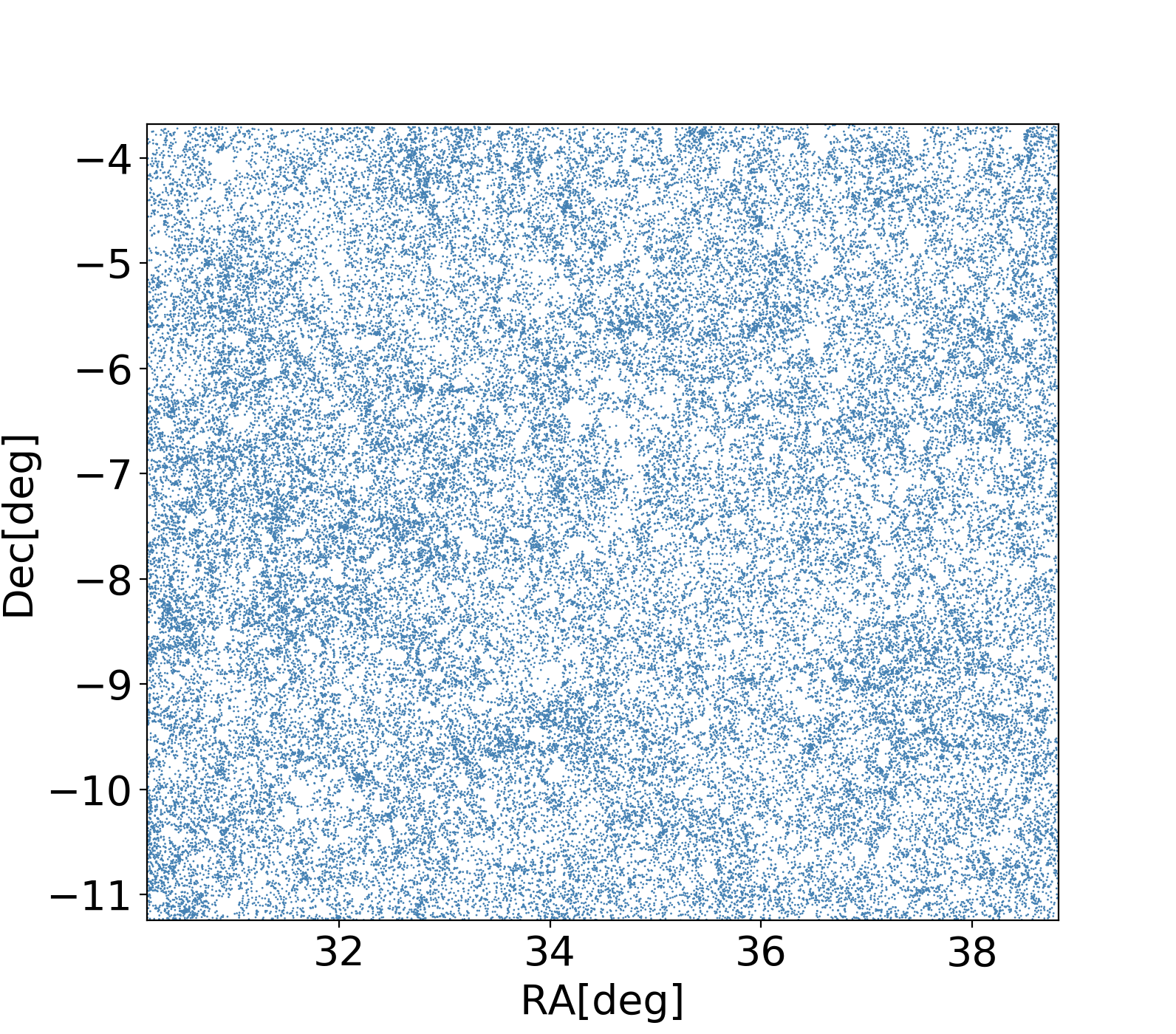}}
   \caption{The panel shows galaxy distribution in W1 with absolute
     magnitude $\mathrm{Mag_i<-21.5}$, $\mathrm{0.335<z<0.535}$.  }
    \label{fig:W1-galaxy}
\end{figure}

\subsection{Observational Data}
\label{section1}

We use the shear catalogue from CFHTLenS (Canada-France-Hawaii Telescope
Lensing Survey)\footnote{\url{http://cfhtlens.org}}, 
which comprises 171 pointings with an effective survey
area of about 154 $\mathrm{deg^2}$. The CFHTLenS data set is based on
the Wide part of CFHTLS carried out in four patches: W1,W2,W3,W4. 
It has deep photometry in five broad bands
$\mathrm{u^* g' r' i' z'}$ (also the $\rm{y'}$ band as a supplement to
$\mathrm{i'}$ band) and limiting magnitude in the $\rm{i'}$ band
of $\mathrm{i'_{AB} \sim 25.5}$. 
 \cite{2012MNRAS.427..146H} presents the CFHTLenS data analysis pipeline, 
 which summarizes the weak lensing data processing with THELI
\citep{2013MNRAS.433.2545E}, shear measurement with Lensfit
\citep{2013MNRAS.429.2858M}, and photometric redshift measurement with
 the Bayesian photometric redshift code \citep{2012MNRAS.421.2355H}.

For each galaxy in the shear catalogue, we are provided with an
inverse-variance weight w, the shape measurement $\mathrm{\epsilon_{1,2}}$, the 
shear correction terms from calibration, the apparent and absolute magnitudes (including
 extinctions and Magnitude error) in five band, the probability distribution function (PDF) 
 of redshift, as well as the peak $\mathrm{z_p}$ of the PDF. The stacked shears are 
 calculated as \citep{2013MNRAS.429.2858M}:
\begin {equation}
\label{EQ-mc}
    \mathrm{\gamma_{1,2}= \frac{\sum_iw_i(\epsilon_{i(1,2)}-c_{i(1,2)})}{\sum_iw_i(1+m_i)}},
\end{equation}
where $\rm{m_i}$ and $\rm{c_i}$ are the multiplicative and additive calibration terms respectively.
    
In order to generate the positions of the LDPs, we use the foreground bright 
galaxies above a certain absolute magnitude, so that the galaxy sample is 
complete in the unmasked areas. This allows us to draw correspondence 
between the observed galaxies and the halos in simulations, and to construct 
the average excess surface density profile around the LDPs in both cases for
comparison. For example, Fig.\ref{fig:w1-full} shows the source distribution
in W1 from CFHTLenS catalogue with i' band apparent magnitude less than 
25.5 in the W1 area. The differences in number densities across different fields
is quite obvious. The empty areas in this map are masked out for bright stars. 
If only the bright galaxies are kept, the sample becomes statistically homogeneous 
in the unmasked areas for redshifts that we are interested in. As an example, 
we show the distribution of galaxies with $\mathrm{0.335<z<0.535}$
and i'-band absolute magnitude $\mathrm{Mag_i\le -21.5}$ in
Fig.\ref{fig:W1-galaxy}.

The positions of the LDPs are identified through the following
procedures:\\

\noindent \emph{\small 1. Generating the LDP candidates}\\

First of all, we require each foreground galaxy to be brighter than a critical
absolute magnitude $\mathrm{Mag_c}$ in the $\rm{i'}$-band, with redshift
between [$\mathrm{z_m}$-0.1, $\mathrm{z_m}$+0.1], where $\rm{z_m}$ 
is the median redshift. We circle around each foreground galaxy with
radius $\mathrm{R_s}$.  Regions within the radius are removed, and
the remaining positions are the candidates for LDPs.  In principle, there are
infinite LDPs. In this work, for simplicity, we put the LDPs on a uniform grid,
with the grid size equal to 0.37 arcmin. The thickness of
the redshift slice is set to 0.2 considering the typical redshift
dispersion $\sigma_z$\footnote{We fit each galaxy redshift PDF with a
  gaussian form to gain the redshift dispersion $\mathrm{\sigma_z}$,
  which is typically around $\sim$0.1.}.  $\mathrm{R_s}$\footnote{
There are no restrictions on the way to remove galaxies, and a lot of tricks can 
be done in this process. For example, each galaxy can be assigned 
with an individual $\rm{R_s}$, according to its luminosity or mass. 
It is also a reasonable way to fix $\rm{R_s}$ for all the galaxies, as done here.}
here is set to 1 or 1.5 arcmin in order to generate enough LDPs.
The $\mathrm{Mag_c}$ here refers to the critical magnitude in one
particular LCDM model (CW1 in \S\ref{section2}).  For other
cosmologies, we use almost the same foreground galaxy
sample\footnote{We always select the same amount of brightest galaxies as the
  foreground galaxies. For different cosmologies, the rank of
  the galaxy brightness may change. For example, for a given apparent
  magnitude at redshift $\mathrm{z_1}$ and $\mathrm{z_2}$, it is
  possible that the absolute magnitude $\mathrm{Mag(z_1,w_{de1})>}$
  $\mathrm{Mag(z_2,w_{de1})}$ but $\mathrm{Mag(z_1,w_{de2})<}$
  $\mathrm{Mag(z_2,w_{de2})}$, due to the change of the
  distance-redshift relation. However, this rarely happens for cosmologies we
  use.}.  To ensure a clean and complete foreground galaxy sample for
generating LDPs, we make three constraints here:

a) We only use sources with $\rm{star\_flag=0}$ to decrease the star
contaminations(but not vanished).  Overall, the fraction of sources with
$\rm{star\_flag=1}$ is around $\rm{3\%}$.  The ratio
becomes $\sim20\%$ for galaxies satisfying $\rm{0.335<z<0.535}$ and
$\rm{Mag_i<-21}$.  In general, the ratio changes with
magnitude and redshift.

b) Galaxies which have two or more close height peaks in redshift PDF
are removed to reduce the redshift uncertainty. Most of these sources
are actually stars.  It further remove $\sim3\%$ sources for galaxies
under condition $\rm{star\_flag}=0$, $\rm{Mag_i<}$-21,
$\rm{0.335<z<0.535}$.
This ratio changes with magnitude and $\rm{z_m}$.

c) We also remove galaxies with absolute magnitude $\rm{Mag_i}<$-99 in
the original catalogues, which indicates problems in the measurement.
This corresponds to the removal of 10 percent additional sources.  In
\S\ref{discussion}, we show that these sources generally yield low
galaxy-galaxy lensing signals, therefore should not be the foreground
galaxies we consider.

After the above selection of galaxies, our i‘-band luminosity function 
is consistent with the CFHTLS-DEEP-SURVEY luminosity function
derived by \cite{2011AJ....142...41R}.\\

\noindent \emph{\small 2. Generating the mask maps}\\

We use the CFHTLenS Mosaic mask files\citep{2013MNRAS.433.2545E} in this step. 
In order to produce the mask maps in (ra, dec) units, we generate the uniform grids for W1-4 firstly. 
Then we apply VENICE\footnote{\url{https://github.com/jcoupon/venice}} with these official files to 
accurately mask these positions near the mask boundaries. \\



\noindent \emph{\small 3. Finding out the LDPs from candidates}\\

Some of the candidate LDPs generated in step 1 should be removed
if they are close to the masked regions.  We require the ratio of the 
masked area to $\rm{\pi R^2_s}$ around each candidate LDP to be less
than 10 percent. To get rid of 
the survey edge effects, we also remove the LDPs whose distances from 
the edges of the survey area are less than $\rm{R_s}$.

For LDPs generated through steps 1 to 3, we measure their stacked
excess surface density $\mathrm{\Delta\Sigma(R)}$ using background
galaxies, and compare it with predictions of different cosmological
simulations introduced in the next section.

\subsection{Simulation}
\label{section2}

We run two sets of simulations named as CW (standing for constant w)
and WZ (referring to w as a function of z), the parameters of which
are given in table \ref{Table1}.  In all of our simulations, we set
$\rm{\Omega_{de}+\Omega_c+\Omega_b=1}$. 2LPT and Gadget2 are used to create
initial conditions and run the simulations \citep{2002MNRAS.333..649S,
  2005MNRAS.364.1105S}. Both CW and WZ simulations are run from
initial redshift 72 with particle number $1024^3$ and boxsize 600
Mpc/h.  We use the FoF group finder to find out the halos, and the
subhalo finder HBT \citep{2012MNRAS.427.1651H} to find the subhalos.

\expandafter{\romannumeral1}) For CW1, we produce the initial
condition following parameters
($\mathrm{\sigma_8=0.85}$,$\mathrm{\Omega_c=0.223}$,
$\mathrm{\Omega_b=0.045,n_s=1}$). For CW2,3,4 the same initial
conditions are used, with updated H(z) for different $\rm{w_{de}}$
model in Gadget2.  The value of $\mathrm{\sigma_8}$ in the 4 simulations reduces
with increasing w.

\expandafter{\romannumeral2}) For the second set, we adopt the best
fit cosmological parameters from \cite{2017NatAs...1..627Z} for
$\mathrm{\Lambda}$CDM and dynamical dark energy w(z) model, and use
CAMB \citep{Lewis:1999bs} to generate the initial power spectrum for
the simulation.

\begin{table}
    \footnotesize
    \centering
     \caption{Simulation parameters.
    \label{Table1} }
    \begin{tabular}{c c c c c c c}
        \hline
        \hline
        Simulation  &$\mathrm{w_{de}}$      &$\sigma_8$  &$\Omega_c$  &$\Omega_b$   &h       &$\rm{n_s}$    \\
        \hline 
        CW1         &-1     &0.85        &0.223       &0.045        &0.71    &1\\
        CW2         &-0.5   &(0.633)          &0.223       &0.045        &0.71    &1\\
        CW3         &-0.8   &(0.789)          &0.223       &0.045        &0.71    &1\\
        CW4         &-1.2   &(0.893)          &0.223       &0.045        &0.71    &1\\
        \hline
        \hline
        Simulation  &$\mathrm{w_{de}}$      &$\mathrm{A_s}$       &$\Omega_c$  &$\Omega_b$   &h       &$\rm{n_s}$\\
        WZ1         &-1     &2.2e-9      &0.2568      &0.0485       &0.679   &0.968\\
        WZ2         &w(z)   & 2.2e-9     &0.24188     &0.04525      &0.702   &0.966\\     
        \hline
    \end{tabular}
\end{table}

 \begin{figure}
    \centering
    \subfigure{
    \includegraphics[width=0.92\linewidth, clip]{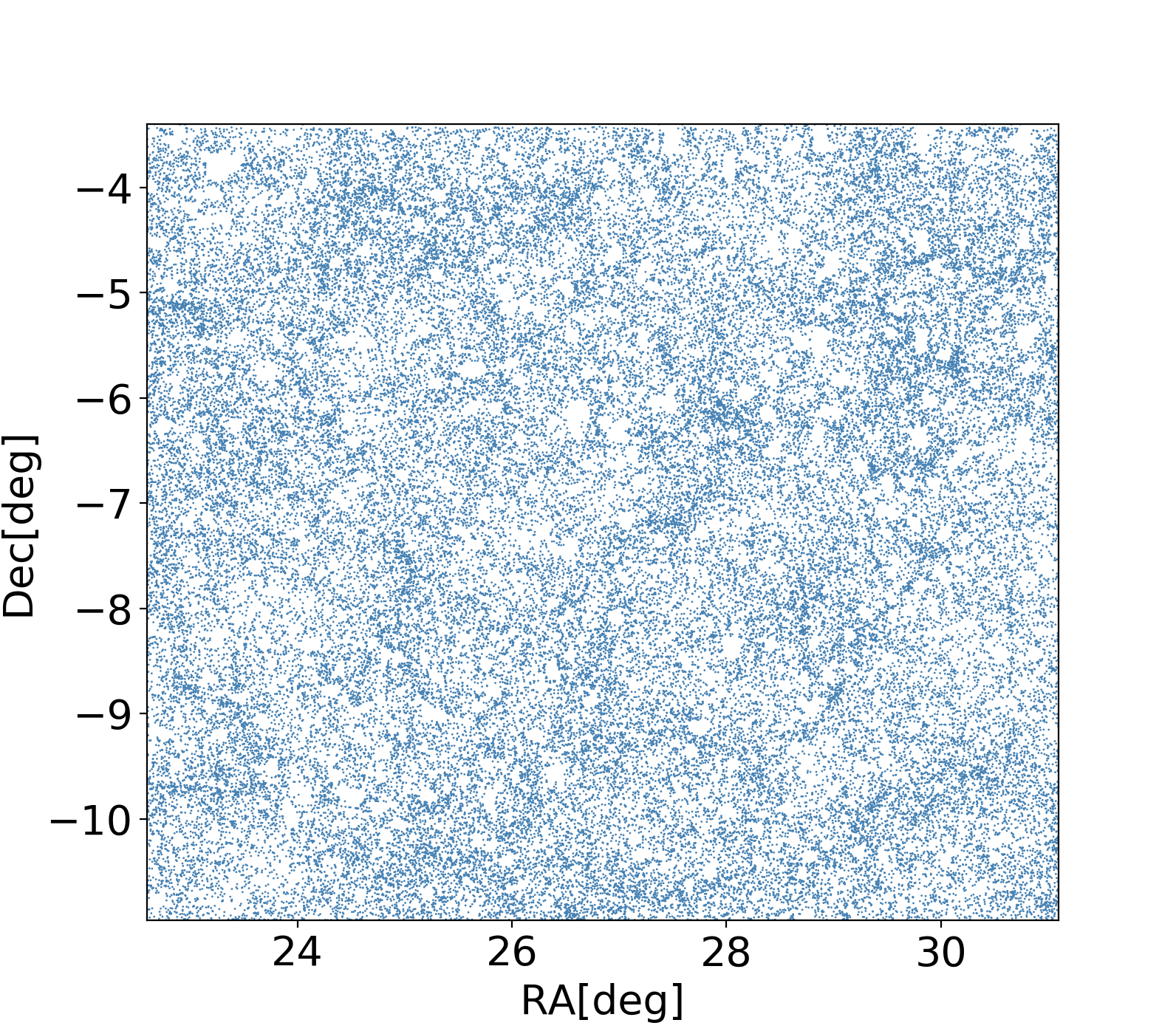}}
  \caption{The figure shows galaxy distribution from one sub-area of
    CW1 simulation for $\mathrm{Mag_i<}$-21.5,
    $\rm{0.335<z<0.535}$. }
    \label{fig:halo-galaxy}
\end{figure}

\begin{figure}
    \centering
    \subfigure{
    \includegraphics[width=0.88\linewidth, clip]{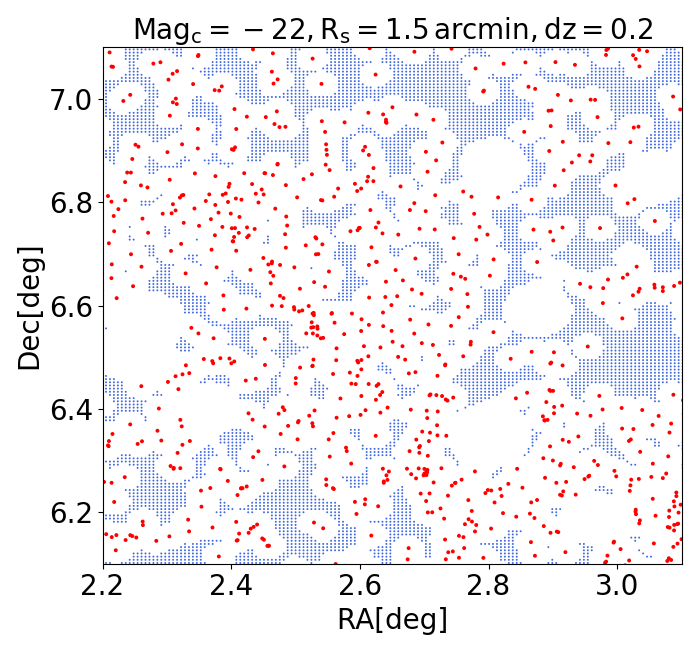}}
  \caption{An example showing the LDP positions (in blue) in our CW1 simulation with $\rm{Mag_c}$=-22, $\rm{R_s}$=1 arcmin,
                $\rm{z_m=0.435}$ and the slice thickness dz=0.2. Red points are the locations of the mock galaxies.  The white regions
                 are either the masked areas or the neighborhood of the mock galaxies.}
    \label{fig:void-halo2}
\end{figure} 


\begin{figure}
    \centering
    \subfigure{
    \includegraphics[width=0.9\linewidth, clip]{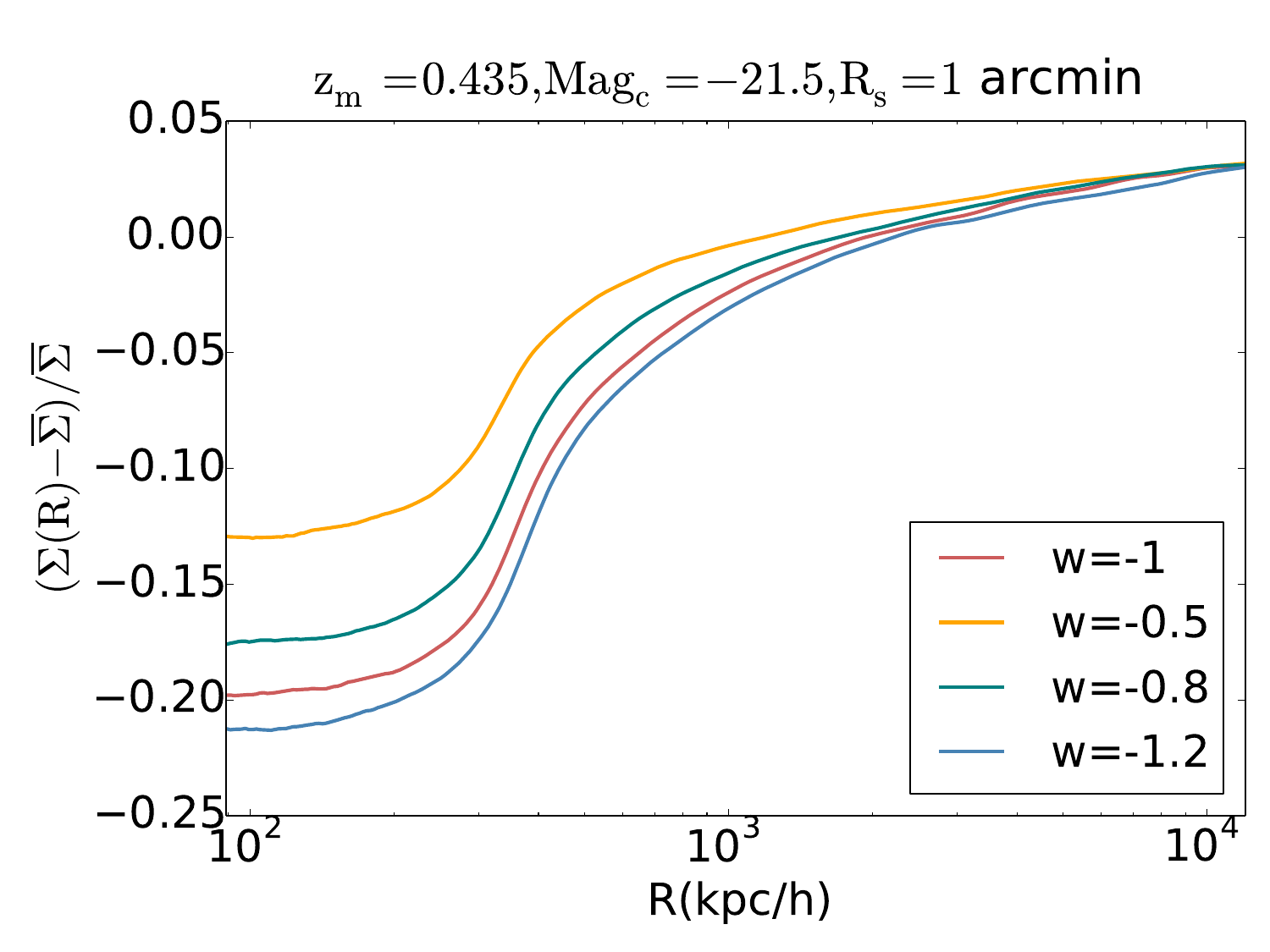}}
  \caption{The curve shows the normalized surface density $\rm{(\Sigma(R)-\overline{\Sigma})/\overline{\Sigma}}$ around
    LDPs defined with $\rm{Mag_c=-21.5},\, \rm{R_s=}$1 arcmin,  $\rm{z_m=}$0.435 and dz=0.2 for
    CW1,2,3,4, where $\rm{\overline{\Sigma}}$ is the mean surface density around the random points.}
    \label{fig:density}
\end{figure}

The LDPs in simulations are defined in the following way:  \\
    
\noindent \emph{\small 1. Connecting halos/subhalos with galaxies
   through SHAM}\\
 
There are different methods in literature to populate galaxies in
 dark matter halos/subbhalos: either through the halo occupation
 distribution and conditional luminosity function models
 \citep{1998ApJ...494....1J, 2002ApJ...575..587B, 2003MNRAS.339.1057Y,
   2005ApJ...633..791Z, 2007MNRAS.376..841V, 2011ApJ...736...59Z,
   2011ApJ...738...45L, 2012ApJ...752...41Y, 2015ApJ...799..130R,
   2016MNRAS.457.4360Z, 2016MNRAS.459.3040G, 2017MNRAS.470..651R,
   2018ApJ...858...30G}, or via subhalo abundance matching processes
 \citep[e.g.][]{2004MNRAS.353..189V, 2006ApJ...647..201C,
   2006MNRAS.371.1173V, 2009ApJ...699..486C, 2010ApJ...717..379B,
   2010MNRAS.404.1111G, 2012MNRAS.423.3458S, 2013MNRAS.433..659H,
   2014MNRAS.437.3228G, 2016MNRAS.460.3100C, 2018arXiv180403097W,
   2018ApJ...860...30Y}. In this work, for simplicity, we use the SHAM
 method to generate mock galaxy catalogs.

 To make the galaxy-subhalo abundance match, we first put the
 simulation box along the line of sight at redshift
 $\mathrm{z_m}$. Then we use galaxies in the redshift range of 
 [$\mathrm{z_m-0.1,z_m+0.1}$] to match the halos/subhalos in the simulation.
 We adopt the masks of observation in the simulation in order to mimic
 the mask distribution in observation. Assuming brighter galaxies are
 formed in bigger halos/subhalos, the connection between the
 halos/subhalos and galaxies are built up by comparing the number of
 halos/subhalos with mass greater than M to the number of galaxies with
 luminosity greater than L:
\begin{equation}
\mathrm{\int_L^{\infty}\phi(L)dL=\int_M^{\infty}[n_h(M)+n_{sh}(M)]dM},
\end{equation}
where
\begin{eqnarray}
       \mathrm{M=}\left\{\begin{array}{ll}
       \mathrm{ M_{acc}}, & \text{~subhalos} \\ 
       \mathrm{ M_{z}}, & \text{~distinct\ halos},
    \end{array}\right.
\end{eqnarray}
where $\rm{n_h/n_{sh}}$ is the number of halos/subhalos, and 
$\mathrm{M_{acc}}$ is the mass at the last epoch when the subhalo is a
distinct halo. It is commonly used in the SHAM models since it is
closely related to the halo merger history. $\mathrm{M_{z}}$ is the
halo mass at redshift z.

 We also add some scatters to the redshifts and luminosities of our
mock galalxies to better mimic the real situation.  Firstly, the
redshift dispersion measured in \S\ref{section1} is added to
halos/subhalos to randomly move their positions in redshift space. We
also update the absolute magnitudes of mock galaxies according to
their redshift errors.  Secondly, we introduce an additional
mass uncertainty of the order $\mathrm{\sigma_{lgM}=0.3}$ to
halos/subbhalos to mimic the dispersion in the galaxy-halo/subhalo
relation.  We find that, the final results are not sensitive to the 
value of $\rm{\sigma_{lgM}}$. We show an example of the galaxy
distribution in CW1 simulation in Fig.\ref{fig:halo-galaxy}, which
looks quite similar to that shown in Fig.\ref{fig:W1-galaxy} for
real galaxies.
\\

 \noindent \emph{\small2. Generating the LDPs}\\
 
 After we generate mock galaxy catalogues, candidate LDPs are
 generated by excluding positions within the radius $\rm{R_s}$ around
 bright galaxies.  We also require that the ratio of the masked area 
 to $\rm{\pi R_s^2}$ around each LDP to be less than 10 percent. 
 An example is given in Fig.\ref{fig:void-halo2}, in which the mock 
galaxies are marked with red color, and LDPs are marked with the
blue color. The white regions are either the masked areas or the 
neighborhood of the mock galaxies within $\mathrm{R_s=1.5}$ arcmin.
The average excess surface density is calculated by stacking the density
 profile around each LDP.  
 
We show the normalized surface density $\mathrm{(\Sigma(R)-\overline{\Sigma})/\overline{\Sigma}}$
around the LDPs of our four different CW simulations in
Fig.\ref{fig:density}, where $\rm{\overline{\Sigma}}$ is the mean surface density around
the random points. 
The LDPs are defined by mock galaxies with $\rm{Mag_c=-21}$, $\rm{R_s=1}$ arcmin,
 $\rm{z_m=0.435}$ and  the slice thickness dz=0.2. The figure shows that a higher w 
 corresponds to a shallower density profile.


 \begin{figure}
    \centering
    \subfigure{
    \includegraphics[width=0.89\linewidth, clip]{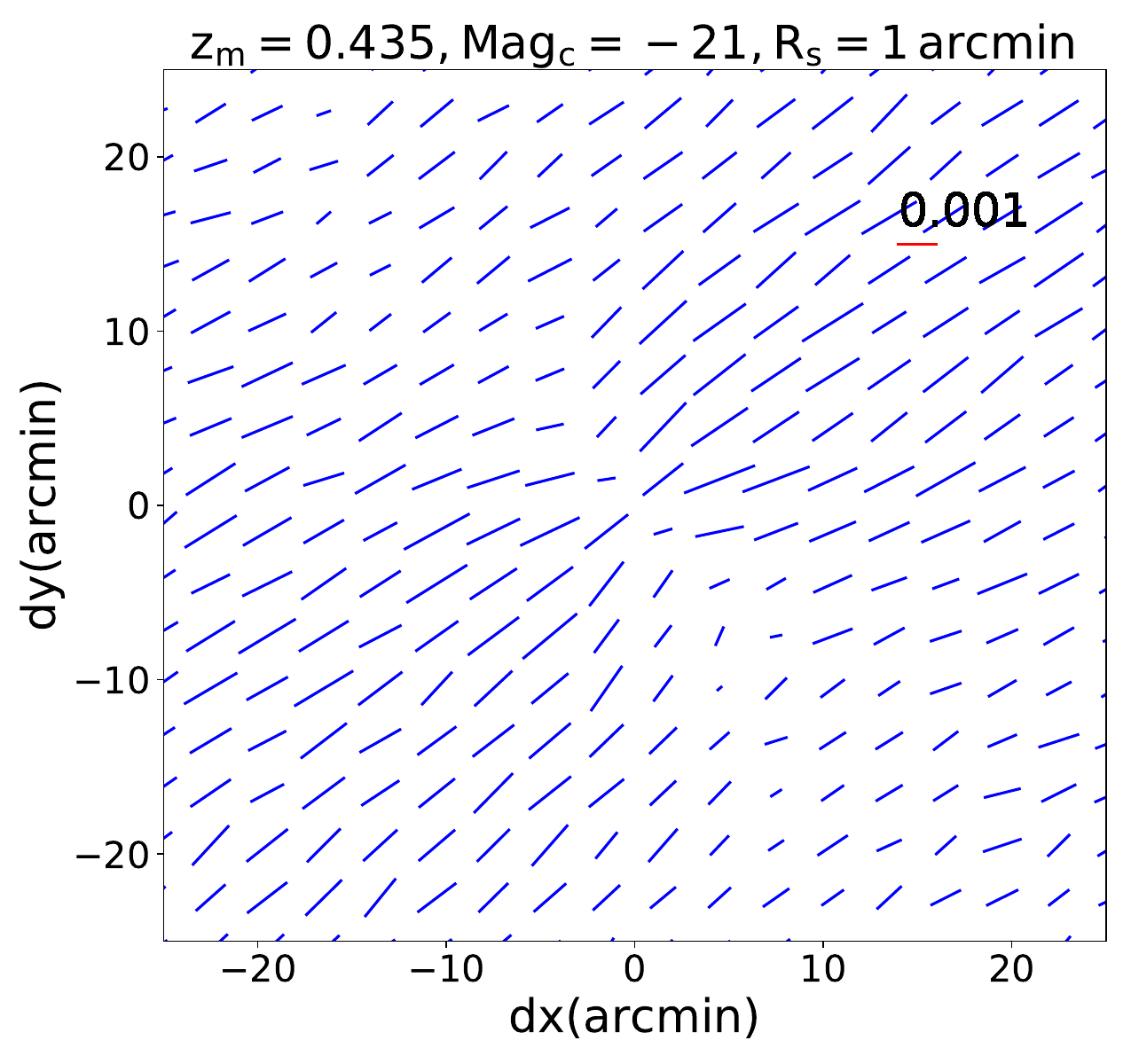}}
    \subfigure{
    \includegraphics[width=0.89\linewidth, clip]{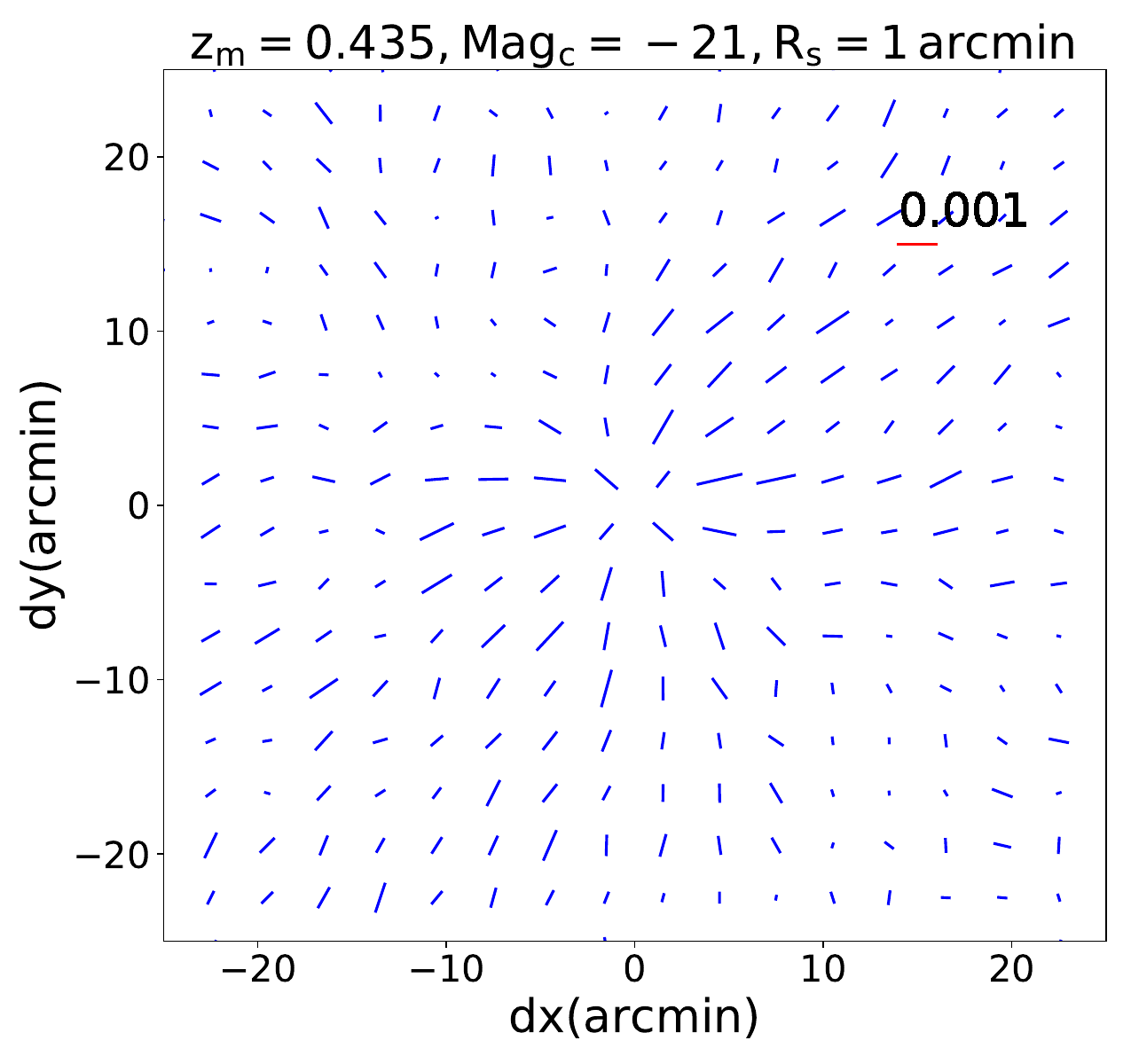}}
  \caption{The 2D stacked lensing signals around the LDPs using the whole CFHTLenS shear catalogue. 
                The upper panel shows the original result, and the lower panel shows results after subtracting 
                the average residual $\rm{\overline{\gamma}_{1,2}}$. The red line represents the shear strength of 0.001.}
    \label{fig-2d-shear}
\end{figure}  


\section{RESULTS}
\label{results}

The 2D stacked shear signals around the LDPs are shown in fig.\ref{fig-2d-shear}.
The upper panel shows the original shear signals calculated with
eq.(\ref{EQ-mc}), in which obvious shear residuals can be found.  
We generate random points in the un-masked regions, whose number is about
ten times that of the LDPs, with the mean shear signals being $\rm{{\overline{\gamma}}_{rand,1,2}\sim3.65\times10^{-4}, 8.5\times 10^{-4}}$.
After subtracting $\rm{\gamma}_{rand,1,2}$ from $\rm{\gamma_{1,2}}$ at the corresponding positions, the shear pattern
corresponding to concave lens emerges, as shown in the lower panel of the figure.
The red lines show the length for shear $\mathrm{\gamma=0.001}$ in the figure. 
In making this figure and the rest of our studies, we remove the background galaxies 
with $\mathrm{\sigma_z>0.2}$ and those with significant multiple peaks in
the redshift PDF to reduce the redshift contamination.  In order to
increase the number of background galaxies and improve the S/N, here
we use galaxies in all fields with shear measurement.  We have also
calculated the 2D shear patterns using only fields that pass the
lensing residual systematics test (shown in \S\ref{discussion}), and
found similar results.  Therefore, the rest of our calculations are
simply based on the full shear catalogue.

More quantitatively, we compare the 1D stacked lensing signals
between observation and simulation.  To estimate the stacked
$\Delta\Sigma(R)$ in observation, we follow the formula (9,10) in
\cite{2015PASJ...67..103N}:
\begin{equation}
\mathrm{ \langle\Delta\Sigma\rangle(R)=\frac{1}{N}
\sum_{a=1}^{N_c}\sum_{s_a: \left|lg\left(\frac{R_{s_a}}{R}\right)\right|<\Delta }w(a,s_a)\Sigma_{cr(a)}\epsilon_{(s_a)+}(R_{s_a})},
\end{equation}
where $\rm{N_c}$ is the number of LDPs, $\rm{2\Delta}$  is the bin size on the logarithmic scale,
$\rm{\epsilon_{(s_a)+}}$ is the tangential ellipticity of the $\rm{s_a}$-th background galaxy for
the a-th LDP, R is the average radius of the background galaxies in that radial bin, 
and N is the normalization factor:
\begin{equation}
\rm{N=\sum^{N_c}_{a=1}\sum_{s_a}w(a,s_a)}\,.
\end{equation}
We only use background galaxies with $\mathrm{z_s \geq z_l+0.1}$ when calculating
 $\mathrm{\Delta\Sigma(z_l,z_s)}$, concerning the fact that both foreground and 
 background galaxies have redshift dispersions.

\begin{figure}
    \centering
    \subfigure{
    \includegraphics[width=0.92\linewidth]{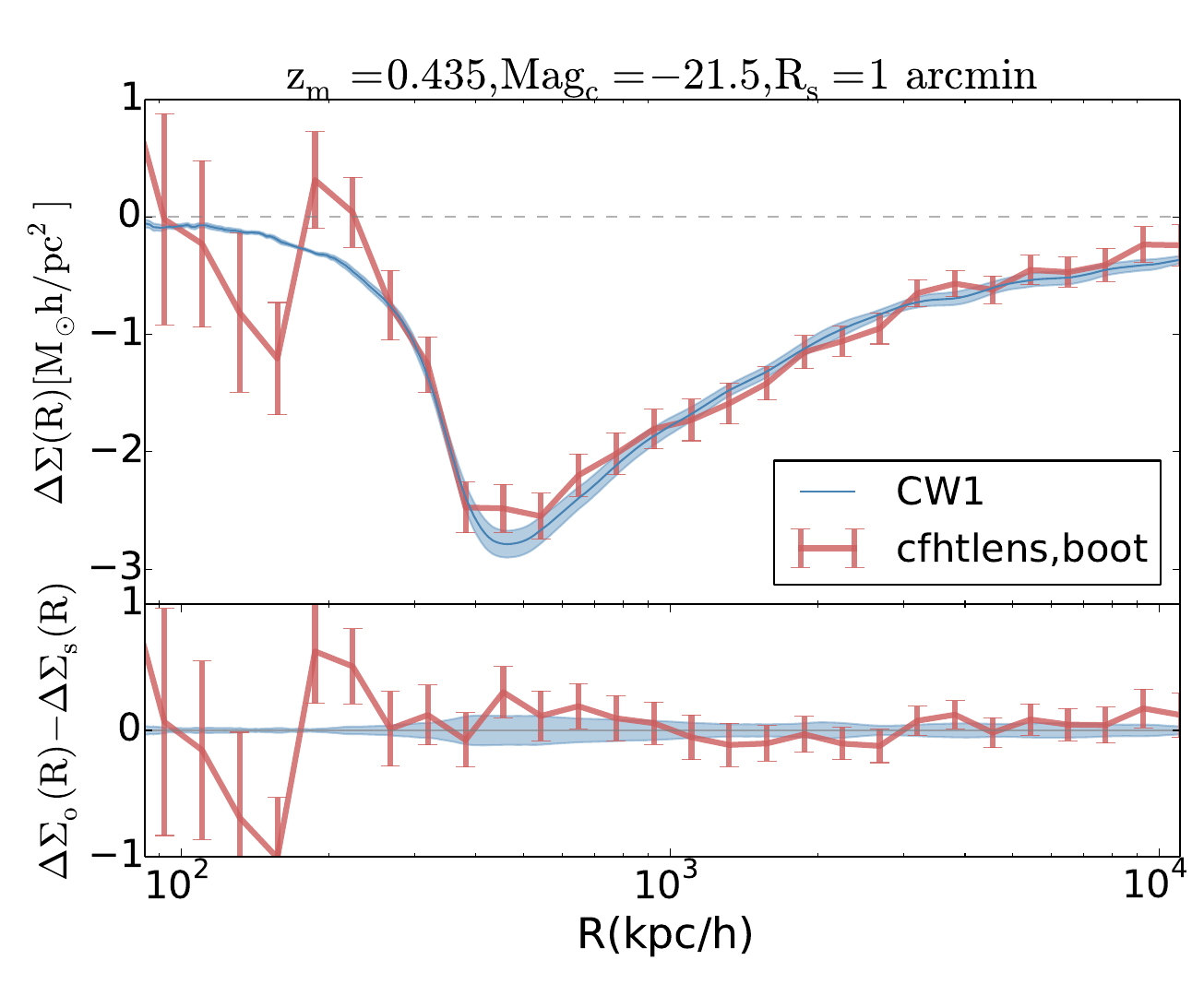}}
  \caption{Results for $\mathrm{z_m=0.435,Mag_c=-21.5,R_s=1}$ arcmin. The blue line is the result from the CW1 simulation. 
                The shaded area shows the size of the cosmic variance estimated with multiple realizations using two CW1 simulations. 
                The red line is the lensing result from the observation, with bootstrap error bars. }
    \label{fig:1panel-0.45}
\end{figure}

\begin{figure}
    \centering
    \subfigure{
    \includegraphics[width=0.98\linewidth]{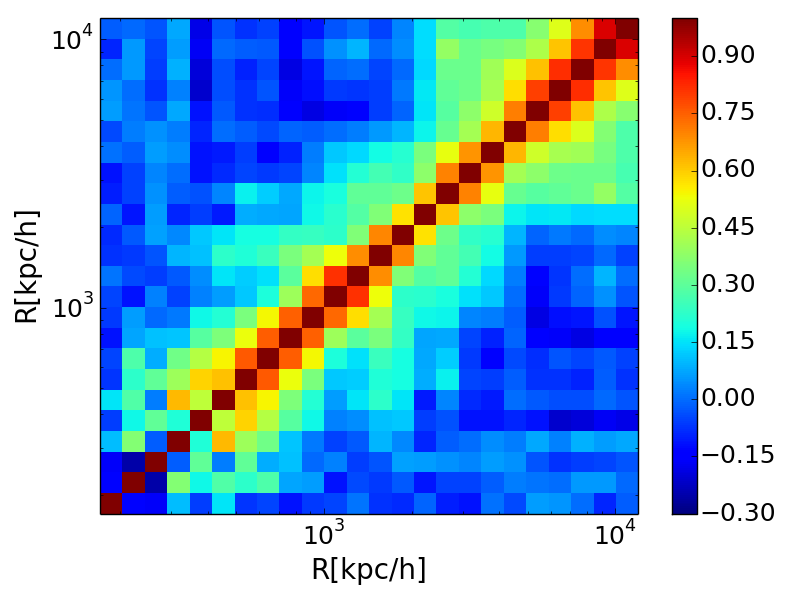}}
    \caption{The normalized covariance matrix for $\mathrm{z_m=0.435,\,Mag_c=-21.5,\,R_s=1\,arcmin}$.}
    \label{fig:cov}
\end{figure}

One of our main results is shown in Fig.\ref{fig:1panel-0.45} with
$\rm{z_m=0.435}$, $\rm{Mag_c=-21.5}$, and $\rm{R_s=}$1 arcmin.  The
red solid line in the upper panel shows $\mathrm{\Delta \Sigma_o(R)}$
calculated with the CFHTLenS catalogues, and the blue solid line shows
$\mathrm{\Delta \Sigma_s(R)}$ from CW1 simulation.  The
$\rm{\Delta \Sigma(R)}$ with subscript ``o" and ``s" represent
observational and simulation signals respectively. The lower panel is
the residual $\mathrm{\Delta \Sigma_o(R)-\Delta \Sigma_s(R)}$, which
shows agreement between the simulation and observation for CW1.  Two
kinds of variances are given in the figure:
(a) We use the simulation of \cite{2007ApJ...657..664J}, which has the
same parameters as CW1, to estimate the cosmic variance for CFHTLenS
survey areas. Each of the two CW1 simulations can be used to generate
4 realizations of about $\rm{150deg^2}$.  The variance estimated from these 
realizations is shown as the shaded area in the figure, with the solid blue line
representing the mean value.   (b) For the observational signals, the variance 
in red line is from bootstrap.  Rather than dividing the foreground space into 
different subregions, we resample LDP groups that are formed by LDPs close 
in space to decrease the impact of masks\footnote{
Given the total area for W1-4 and the largest radius $\rm{R_{max}}$ in Fig.\ref{fig:1panel-0.45}, 
the number of groups is estimated by dividing the total area by $\rm{4R^2_{max}}$. 
The mean number of LDPs within a group is derived by dividing the total number of LDPs
with the number of groups. Then we divide all the LDPs into regularly placed small cells, with each cell 
containing a few of LDPs. The cells are added to the groups one by one. Whenever the
size of a group in a row reach $\rm{2R_{max}}$, we move to another row. If the number of 
LDPs in a group equals to the mean number, the assignment for this group is terminated. 
This procedure is repeated untill the last cell is assigned to a group.  In this way, the LDPs are
divided into groups with similar numbers, avoiding the fluctuations due to the masking effects. }. 
In this way,  the error bar is relatively stable even if the group size is changed 
for several times. To fully capture the 
covariance matrix, the size of the LDP group is set to be two times larger than the largest
radial bin. The corresponding normalized covariance matrix is shown in Fig.\ref{fig:cov}. 
Correlations between the neighboring radial bins can be found in the figure.


 
\begin{figure*}
    \centering
     \subfigure{
    \includegraphics[width=1\linewidth, clip]{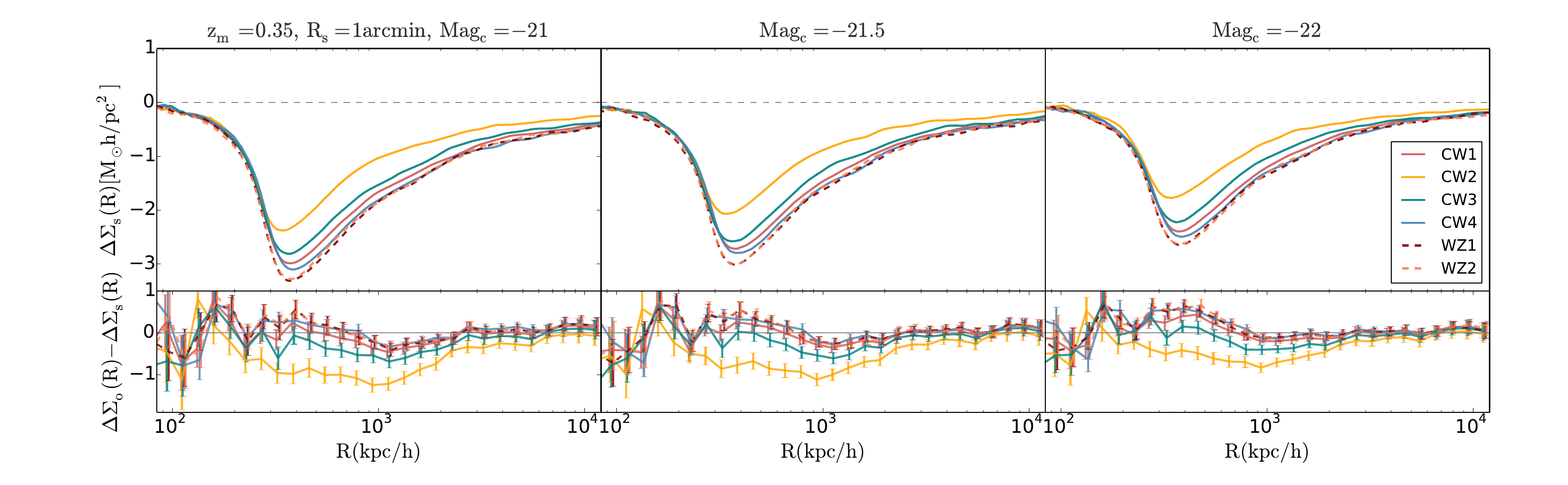}}
    \subfigure{
    \includegraphics[width=1\linewidth, clip]{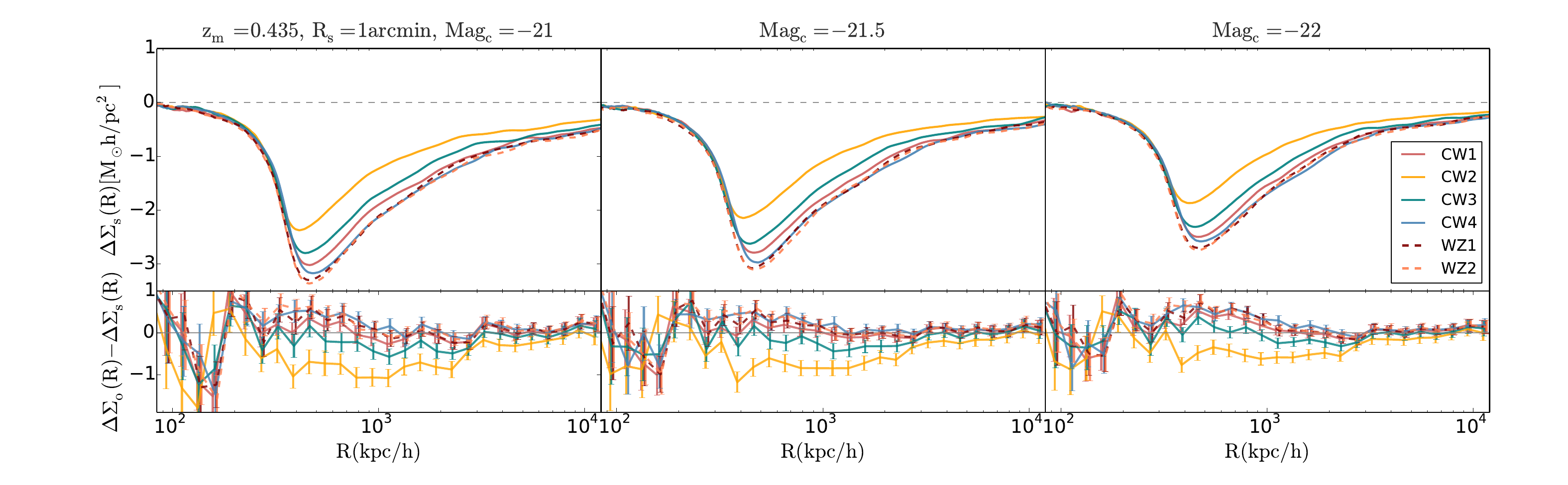}}
    \subfigure{
    \includegraphics[width=1\linewidth, clip]{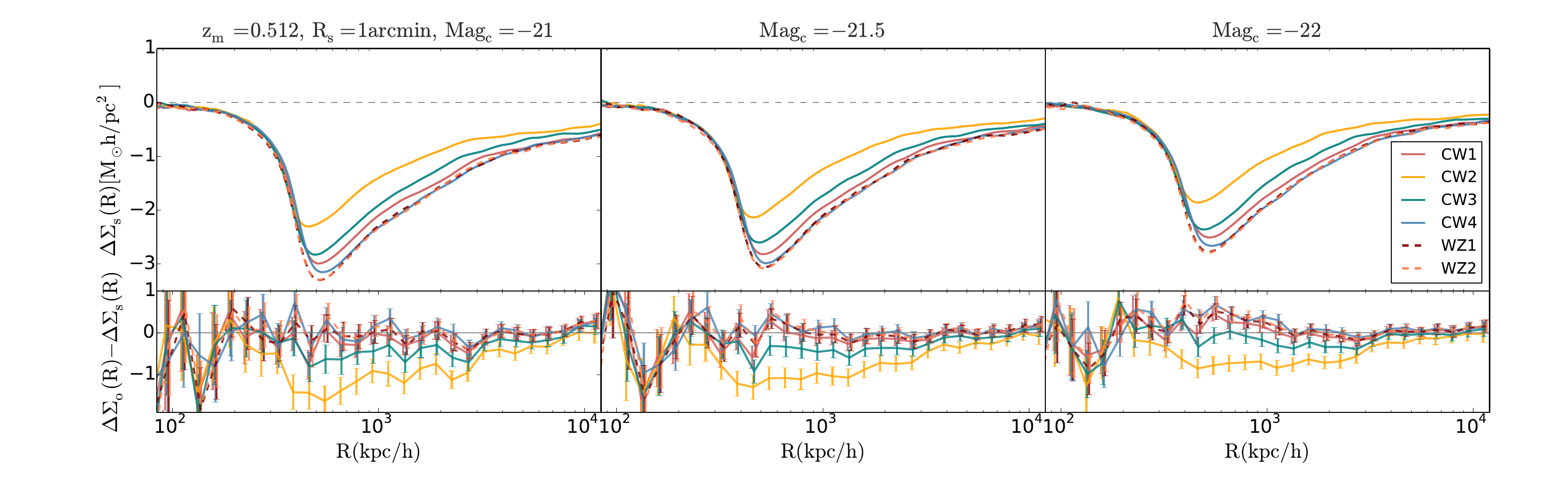}}
    \caption{The average excess surface density profile around LDPs that are defined with $\rm{R_s=1\,arcmin}$ and different 
                  choices for the magnitude cut and redshift range. The left, middle, and right columns are for $\rm{Mag_c=-21,-21.5,-22}$
                  respectively, and the top, middle, and bottom rows for $\rm{z_m=0.35, 0.435, 0.512}$. The upper part of each panel 
                  shows $\rm{\Delta \Sigma_s(R)}$ from simulations. The lower part shows the residuals after subtracting 
                  $\rm{\Delta \Sigma_s(R)}$ from $\rm{\Delta \Sigma_o(R)}$. 
   \label{fig:6simu-1}}
\end{figure*}

\begin{figure*}
    \centering
    \subfigure{
    \includegraphics[width=1\linewidth, clip]{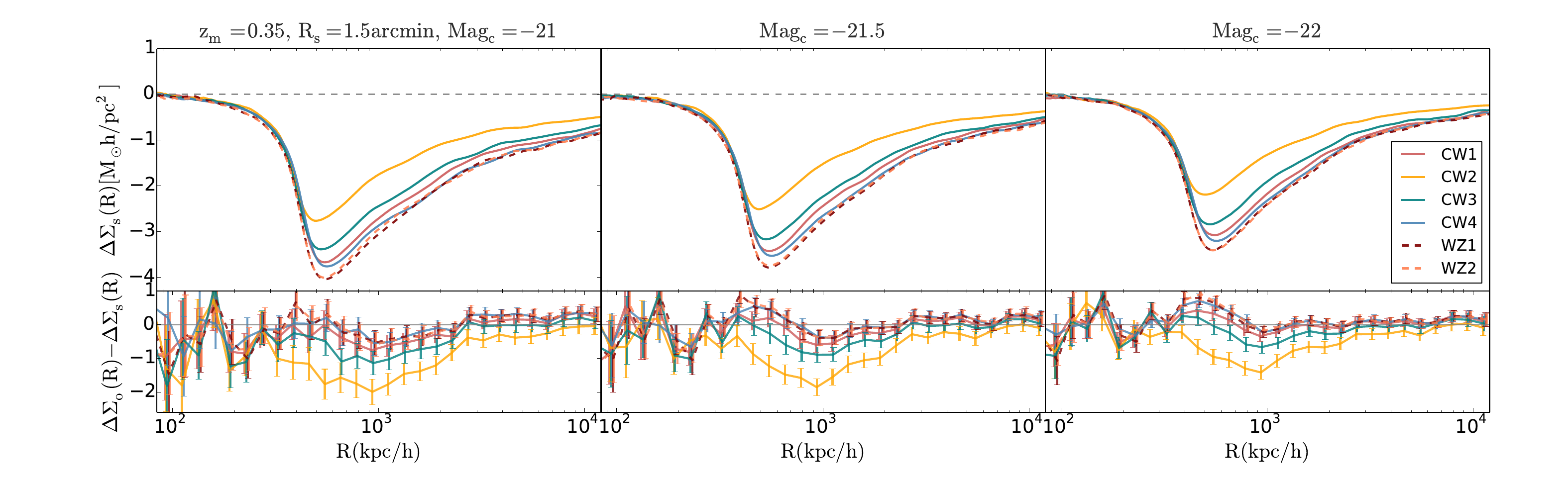}}
    \subfigure{
    \includegraphics[width=1\linewidth, clip]{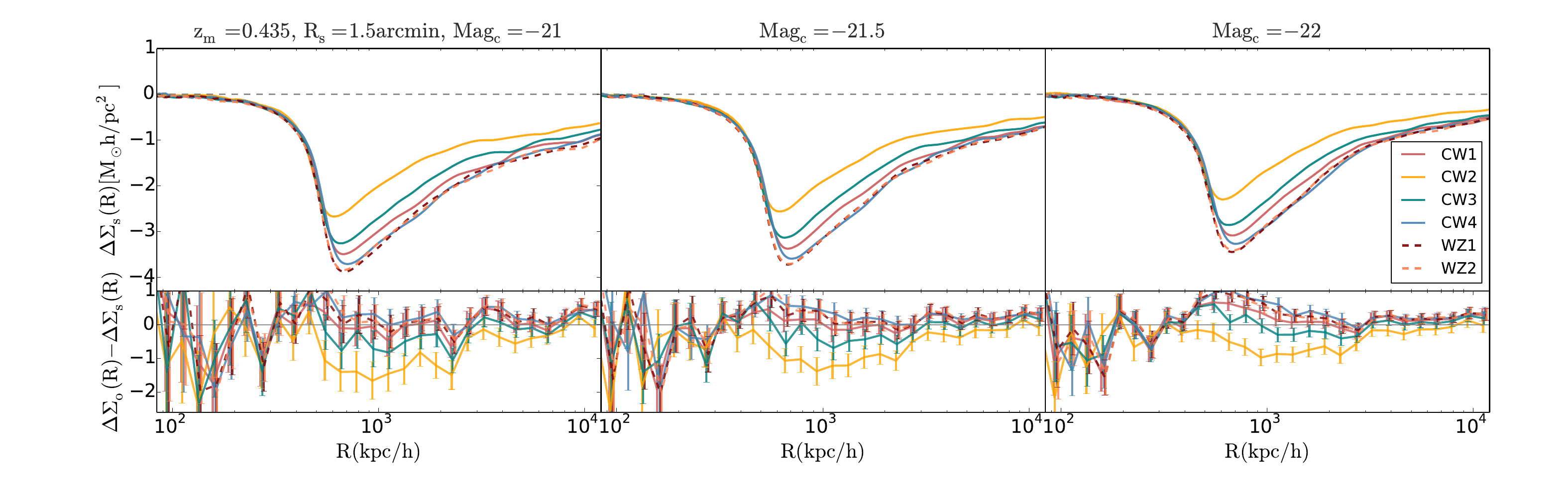}}
    \subfigure{
    \includegraphics[width=1\linewidth, clip]{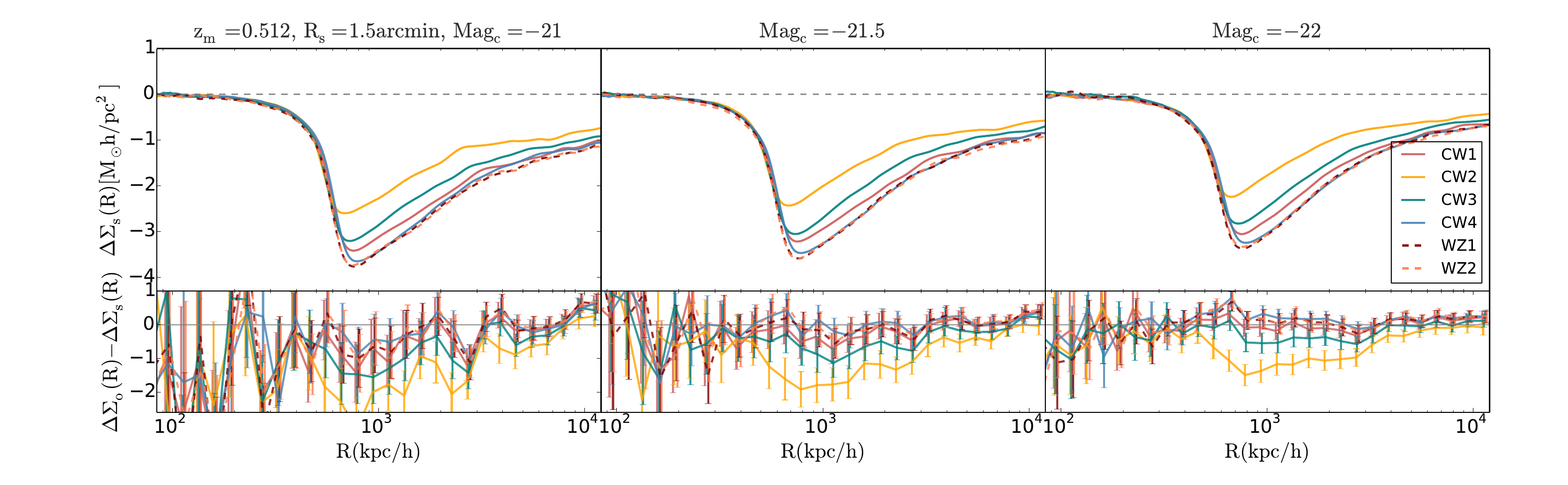}}
    \caption{Similar to fig.\ref{fig:6simu-1}, but with $\rm{R_s=1.5\,arcmin}$.}\label{fig:6simu-1.5}
\end{figure*}

\begin{figure*}
    \subfigure{
    \includegraphics[width=0.93\textwidth, clip]{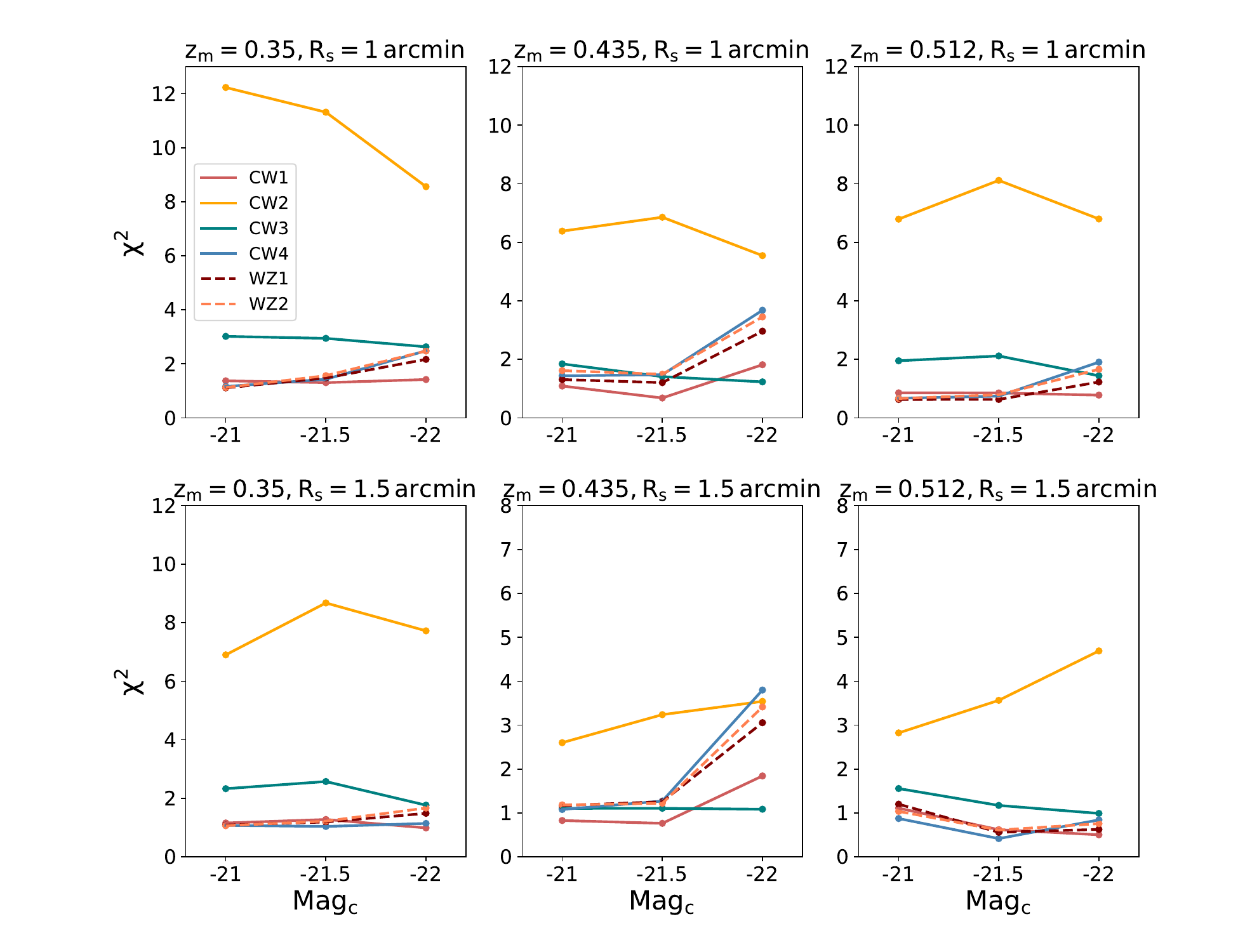}}
    \caption{ The reduced $\mathrm{\chi^2}$ for six cosmologies with different choices of $\mathrm{Mag_c}$, $\mathrm{R_s}$ and $\mathrm{z_m}$. 
    The upper three panels are for $\mathrm{R_s=}$ 1 arcmin, and lower panels are for $\mathrm{R_s=}$ 1.5 arcmin. The left, middle, 
    and right columns are for $z_m=0.35, 0.435, 0.512$ respectively. The horizontal axis in each plot is the magnitude cut 
    $\mathrm{Mag_c}$. \label{fig-kaf}}
\end{figure*}

The total S/N ratio of the observational lensing signal can be
calculated as:
 \begin{eqnarray}
    \label{EQ:SNR}
&&\mathrm{\left(\frac{S}{N}\right)^2=\sum_{i,j}\Delta\Sigma_o(\theta_i)C_{i,j}^{-1}\Delta\Sigma_o(\theta_j)}, \\  \nonumber
&& \rm{C_{i,j}^{-1}=\frac{N_S-N_D-2}{N_S-1}C_{i,j}^{\ast-1}}, \\  \nonumber
&& \rm{C_{i,j}^{\ast}=cov(\Delta\Sigma_o(\theta_i),\Delta\Sigma_o(\theta_j))},
\end{eqnarray}
where the summation runs over all the radial bins. 
Since the sampled precision matrix $\rm{C_{i,j}^{\ast-1}}$ is biased due to the noise in $\rm{C_{i,j}^\ast}$, the
 unbiased estimator $\rm{C_{i,j}^{-1}}$ is taken \citep{2007A&A...464..399H,2013MNRAS.432.1928T}.
Here $\rm{N_S}$ is taken as the number of the LDP groups, and $\rm{N_D}$ is the number of the data bins.
The S/N for the
red line in Fig.\ref{fig:1panel-0.45} is 20.473, which is
significant for us to constrain cosmologies and shows the advantage of
stacking signals around LDPs.

We repeat the procedures for $\mathrm{R_s=1,1.5\,arcmin}$, redshift
bins $\mathrm{z_m=0.35,0.435,0.512}$, and Magnitude cuts
$\mathrm{Mag_c=-21,-21.5,-22}$. It enables us to make multiple
comparisons with limited information. Since the apparent magnitude
limit for the i'-band is higher than 24.5 on average, it allows us to make
the volume-limited samples for three foreground redshifts and three magnitude $\mathrm{Mag_c}$ cuts. 

Fig.\ref{fig:6simu-1} and Fig.\ref{fig:6simu-1.5} show detailed scale
dependent comparisons between simulation and observational results for
$\mathrm{R_s=}$1 arcmin and $\mathrm{R_s=}$1.5 arcmin respectively.
Panels in the horizontal direction of Fig.\ref{fig:6simu-1} and
Fig.\ref{fig:6simu-1.5} are for three choices of $\mathrm{Mag_c}$, and
for three $\mathrm{z_m}$'s in the vertical direction.  The arrangement
of each panel is similar to that of Fig.\ref{fig:1panel-0.45}.  One
can see from the figures that there is not a single best model for all
cases.  It also happens that the best model in one case does not
perform well at all radius scales.  This situation seems to be worse
for $\mathrm{R_s=}$1.5 arcmin. This is likely due to the fact that the
number of LDPs for $\mathrm{R_s=}$1.5 arcmin is much lower than that
of $\mathrm{R_s=}$1 arcmin.
 
In order to compare the results more
directly we introduce  the reduced $\mathrm{\chi^2}$ to describe the discrepancy
between observational and simulation signals:
\begin{eqnarray}
\label{EQ:kaf}
&&\mathrm{\chi^2=\frac{1}{N_{bin}}\sum_{i,j}\delta\Delta\Sigma(\theta_i)C_{i,j}^{-1}\delta\Delta\Sigma(\theta_j)}, \\ \nonumber
&&\mathrm{\delta\Delta\Sigma(\theta_i)=\Delta\Sigma_o(\theta_i)-\Delta\Sigma_s(\theta_i)}.
\end{eqnarray}
When calculating S/N or $\mathrm{\chi^2}$, we do not consider the
cosmic variance.  The $\mathrm{\chi^2}$ results for all cases are shown in                       
Fig.\ref{fig-kaf}.  The upper three panels are for $\mathrm{R_s=}$ 1
arcmin, and lower panels are for $\mathrm{R_s=}$ 1.5 arcmin.  The
left, middle, and right panels are for $\rm{z_m=0.35,0.435}$ and 0.512
respectively.  The horizontal axis is $\mathrm{Mag_c}$.  The four
solid lines in each panel show results for CW1-4 cosmologies, and the
dashed lines are for WZ1-2.  Among six simulations, CW1 and WZ1 are
two different $\rm{\Lambda CDM}$ models.  From these panels we find:
\begin{enumerate}
\item In all the panels, CW2 (w=-0.5) always has the largest $\rm{\chi^2}$
 compared with other cosmologies;
    
\item  CW3 (w=-0.8) has the second largest $\rm{\chi^2}$  in most cases;
    
 \item The other four models, including the two $\mathrm{\Lambda}$CDM models (CW1,WZ1), 
 the CW4 (w=-1.2) model, and WZ2 (dynamical w(z)), all have comparably low $\rm{\chi^2}$ i
 n most cases. The most pronounced exception is in the case of $\rm{z_m=0.435}$, 
 $\rm{R_s=1.5}$ arcmin, and $\rm{Mag_c=-22}$, in which the CW3 model yields the
  lowest $\rm{\chi^2}$ in contrast.
    
\end{enumerate}

\section{CONCLUSION AND DISCUSSIONS}
\label{discussion}
 
    \begin{figure}
    \centering
    \subfigure{
    \includegraphics[width=0.93\linewidth, clip]{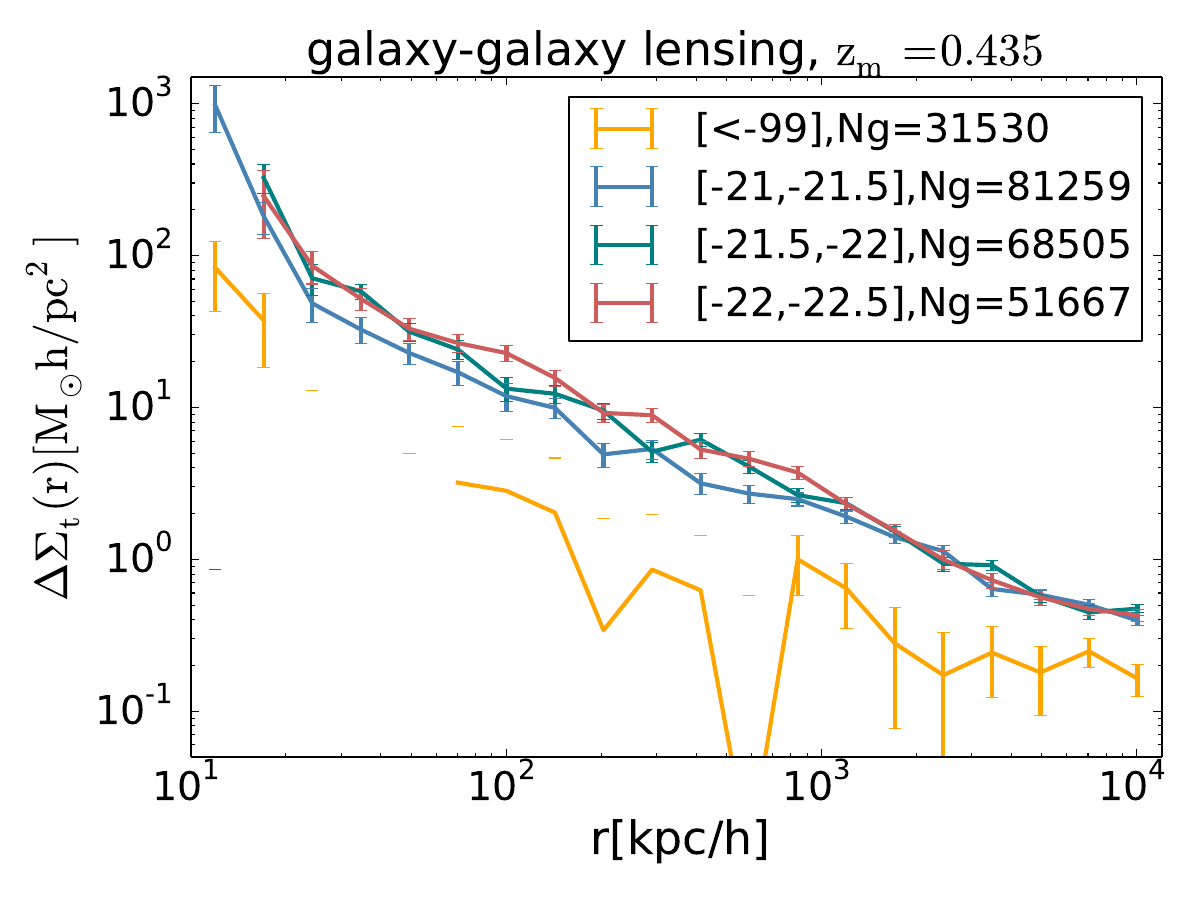}}
    \caption{ \label{fig-gg} Galaxy-galaxy lensing signals for foreground galaxies in the
                  Magnitude bins of [-21,-21.5], [-21.5,-22], [-22,-22.5] and [$\rm{<}-99$], and
                   redshift range of [0.335,0.435]. Ng is the number of foreground galaxies. }
    \end{figure} 
    
    \begin{figure*}
    \centering
    \subfigure{
    \includegraphics[width=0.43\textwidth, clip]{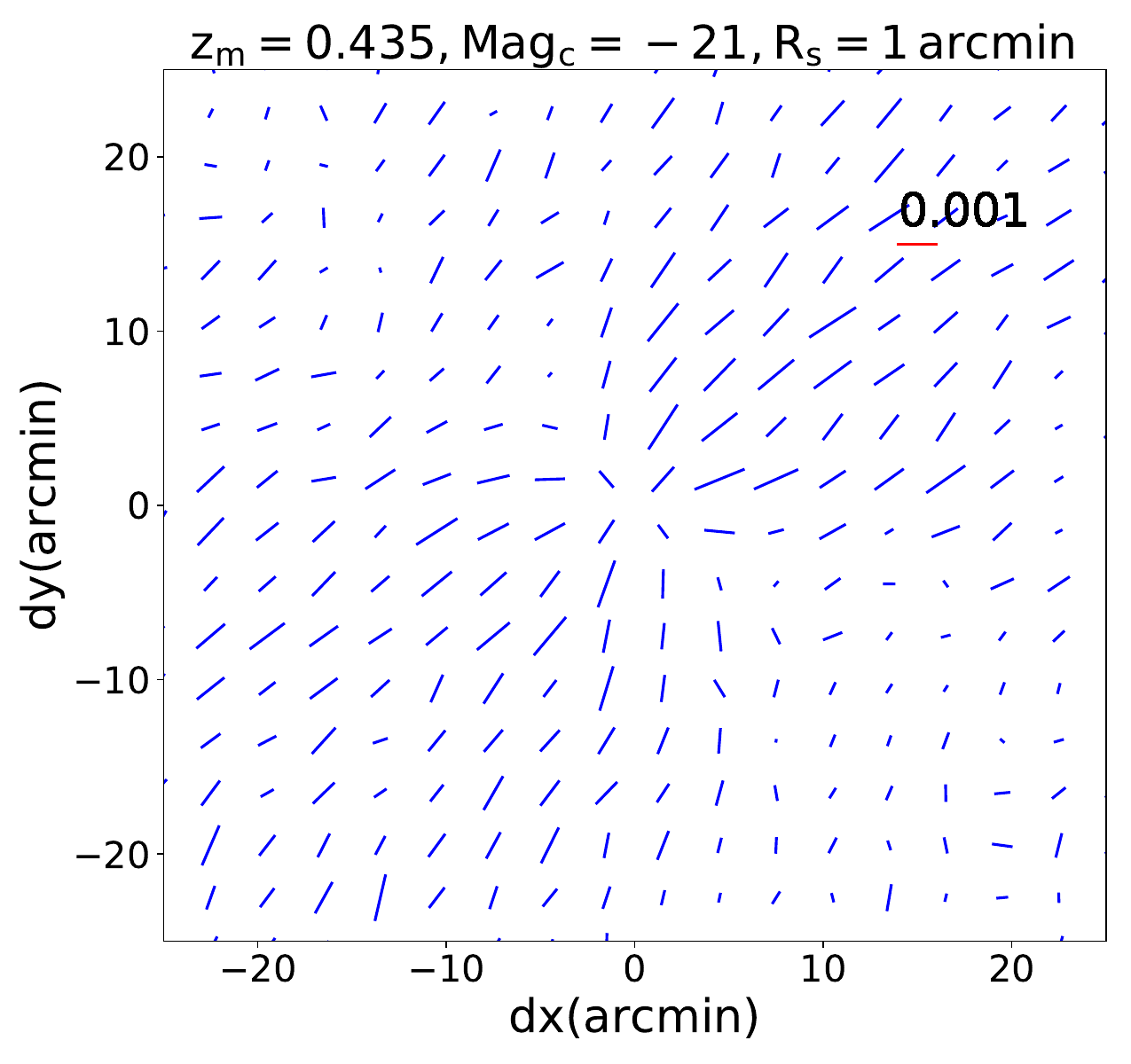}}
    \subfigure{
    \includegraphics[width=0.43\textwidth, clip]{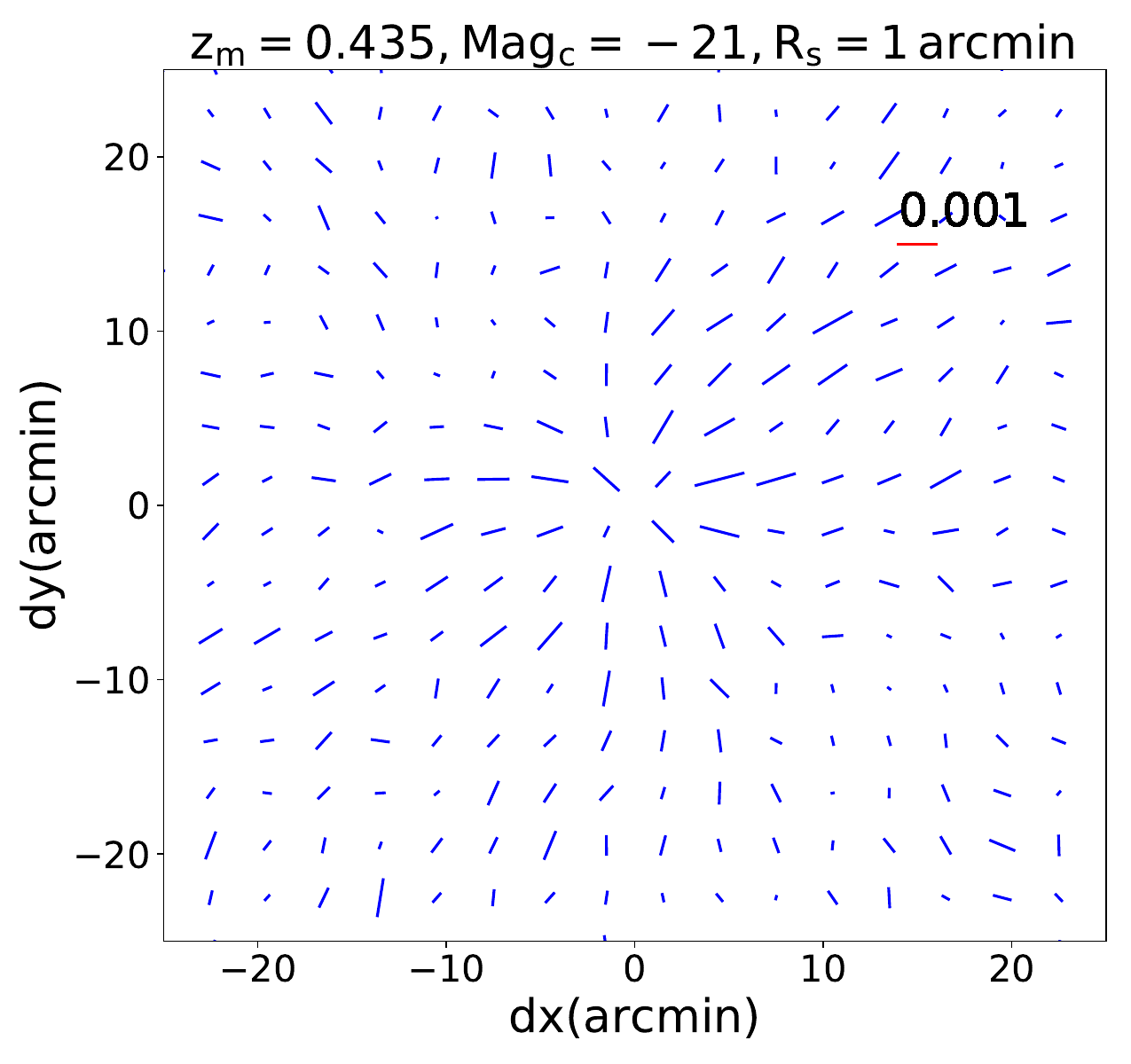}}
    \caption{The 2D stacked lensing signals around the LDPs using only the CFHTLenS fields that pass
                  the lensing residual systematics tests. The left panel shows the original result, and the right 
                  panel shows results after subtracting the average residual $\rm{\overline{\gamma}_{1,2}}$. 
                  The red line represents the shear strength of 0.001. 
    }
    \label{fig-2d-good-shear}
    \end{figure*} 

In this paper we study the stacked lensing signals around the low-density-positions (LDPs), which are defined as places that are devoid of foreground bright galaxies in projection. We show how to define the foreground galaxy population and locate the LDPs in the presence of masks using the CFHTLenS data. Different redshift ranges and magnitude cuts are considered for the foreground population. The measured excess surface density profiles can be compared with the predictions from simulations. The comparison is made available by drawing correspondence between galaxies and halos/subhalos via SHAM.

With the CFHTLenS shear catalogue, we have successfully measured the lensing signals around the LDPs with a high significance. These measurements are used to constrain dark energy models using simulated galaxies that have similar survey selection effects. We run six cosmological simulations [CW(1,2,3,4) and WZ(1,2)] with different dark energy equations of state, for the purpose of reproducing the mean surface density profile around the LDPs in observation. The cosmological parameters of the six simulations are given in table \ref{Table1}. 

Our results of the surface density measurement indicate that the CW2 ($w=-0.5$) and CW3 ($w=-0.8$) models are not favored. The two $\rm{\Lambda CDM}$ models (CW1 and WZ1), as well as the CW4 ($w=-1.2$) and WZ2 ($w(z)$ of \cite{2017NatAs...1..627Z}) models, all achieve reasonably good and similar agreement with the observation. The comparisons are made for three foreground redshift bins, three magnitude cuts, and two critical radii for the definition of the LDPs.   

There are a number of problems that may impact our results. Here we outline some of them:

\subsection{The Impact of Throwing Away Sources with Absolute Magnitude $\rm{Mag_i}<$-99 in the Shear Catalogue}
 
We show the galaxy-galaxy
lensing signals in Fig.\ref{fig-gg} for foreground galaxies in the redshift range of [0.335,0.435]. The blue, green, red and yellow lines are for galaxies with magnitudes in the range of [-21,-21.5], [-21.5,-22] , [-22,-22.5], and
[$\rm{<}$-99] respectively. For galaxies of $\rm{Mag<}$-99, their lensing signals are quite low, indicating that they likely correspond to less massive sources 
on average (or even not galaxies). So we think it is safe for us to remove them from the foreground galaxies when generating the LDPs.

\subsection{The Impact of the Fields Which Do Not Pass the Lensing Residual Systematics Tests}
  
 As described in \citep{2012MNRAS.427..146H}, some CFHTLenS fields do
not pass the lensing residual systematics tests, which should not be
used in shear two-point correlation function. In fig.\ref{fig-2d-good-shear}, we show the 2D stacked shear patterns around the LDPs using only fields that pass the lensing residual systematics tests. The left panel shows the original result, and the right panel shows results after subtracting the random points signals $\rm{\gamma_{rand,1,2}}$. The red line represents the shear strength of 0.001. From the right panel we can find that the shear patterns do not change too much from that shown in the lower panel of fig.\ref{fig-2d-shear}, which uses all the fields. For the 1D signals, the conclusion is similar. We therefore simply use the whole CFHTLenS shear catalogue in this work.

\subsection{The Impact of the $\rm{\sigma_8}$ When Comparing The Lensing Signals.}
\label{new-simu}

The initial conditions for the CW1-4 simulation are fixed as mentioned in \S\ref{section2}, supposing the early-time 
amplitude $\rm{A_s}$ is well constrained by the CMB. However, it may be interesting to ask what if we keep the late-time
amplitudes the same for these simulations, although with very different $\rm{A_s}$. So we run four new simulations here as a
comparison.  Three new simulations are run for the CW set, named as CW5,6,7. Also, one new simulation is run for the 
WZ set, named as WZ3. For CW5,6,7, the parameters ($\rm{\Omega_c,\Omega_b,w_{de},h}$) are set the same as in 
CW2,3,4 respectively, with the $\rm{\sigma_8}$ being the same as that of  CW1. For WZ3, its parameters($\rm{\Omega_c,\Omega_b,w_{de},h}$) 
are identical with WZ2,  with the $\sigma_8$ taken from WZ1. The parameters of the simulations are given in table \ref{Table2}. 
   
\begin{figure*}
    \centering
     \subfigure{
    \includegraphics[width=1\linewidth, clip]{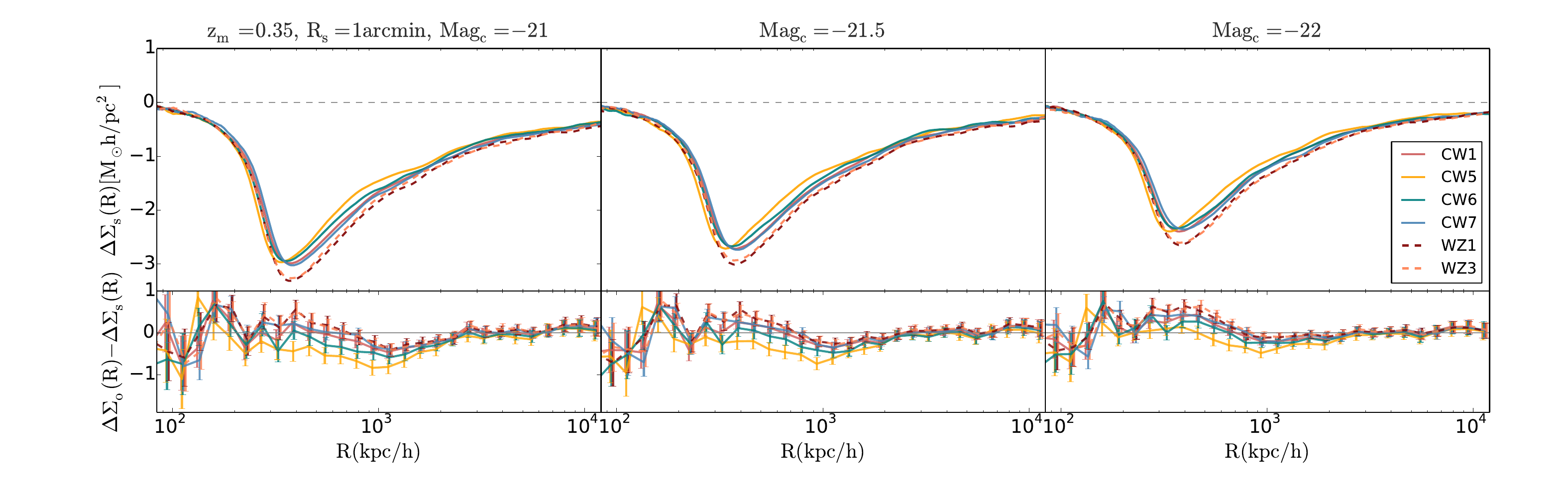}}
    \subfigure{
    \includegraphics[width=1\linewidth, clip]{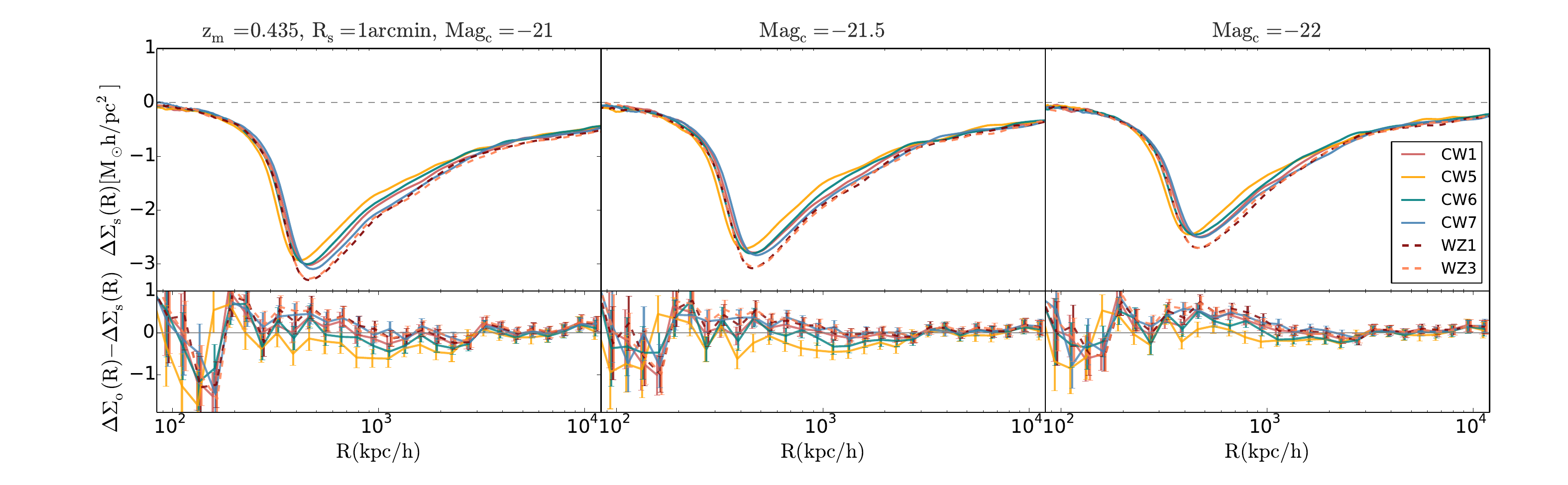}}
    \subfigure{
    \includegraphics[width=1\linewidth, clip]{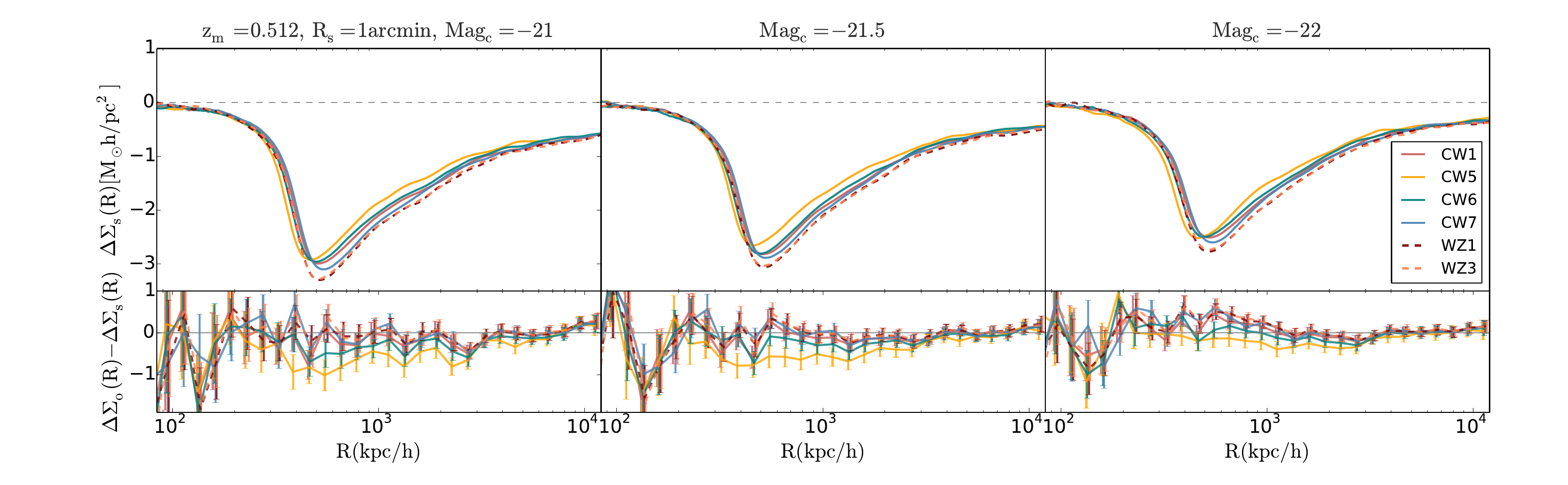}}
    \caption{Similar to Fig.\ref{fig:6simu-1}, but with a new set of simulations described in \S\ref{new-simu}. CW5,6,7 all share the same $\rm{ \sigma_8}$ as that of CW1, and WZ3 has the same $\rm{\sigma_8}$ as that of WZ1.}
   \label{fig:6simu-1-new}
\end{figure*}

\begin{figure*}
    \centering
    \subfigure{
    \includegraphics[width=1\linewidth, clip]{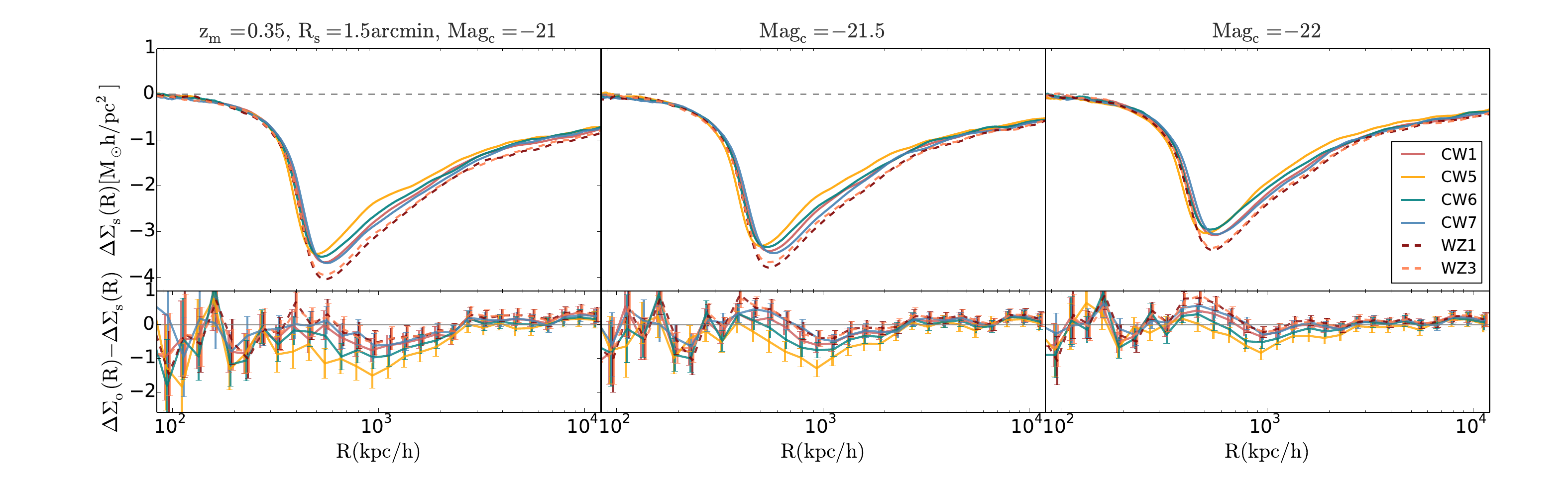}}
    \subfigure{
    \includegraphics[width=1\linewidth, clip]{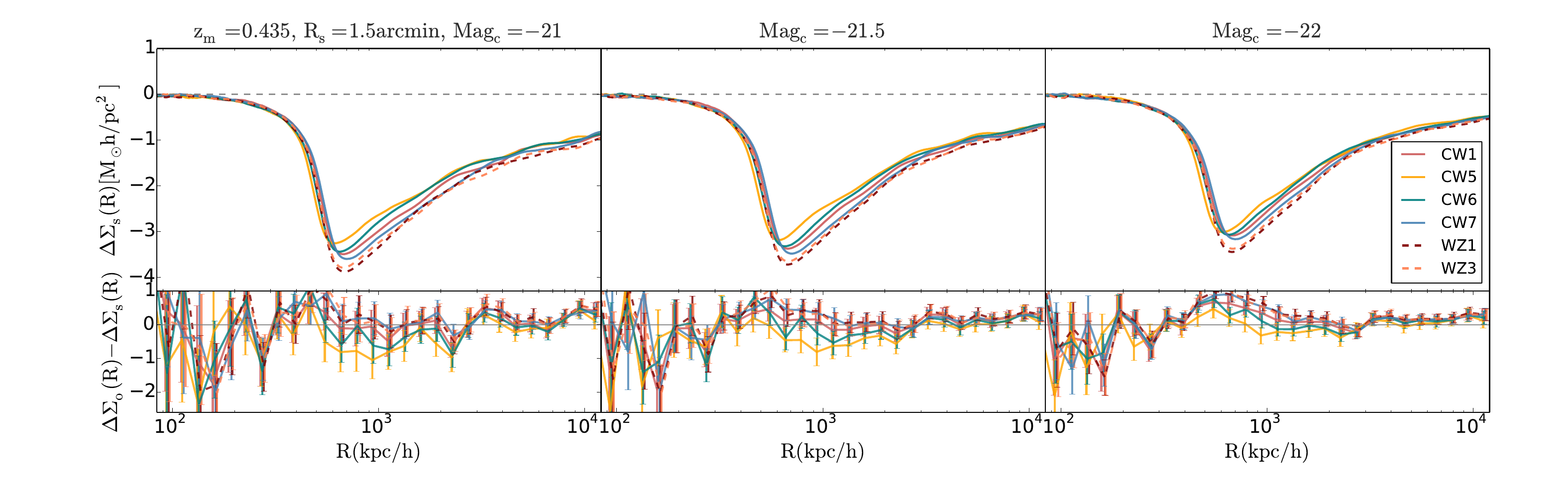}}
    \subfigure{
    \includegraphics[width=1\linewidth, clip]{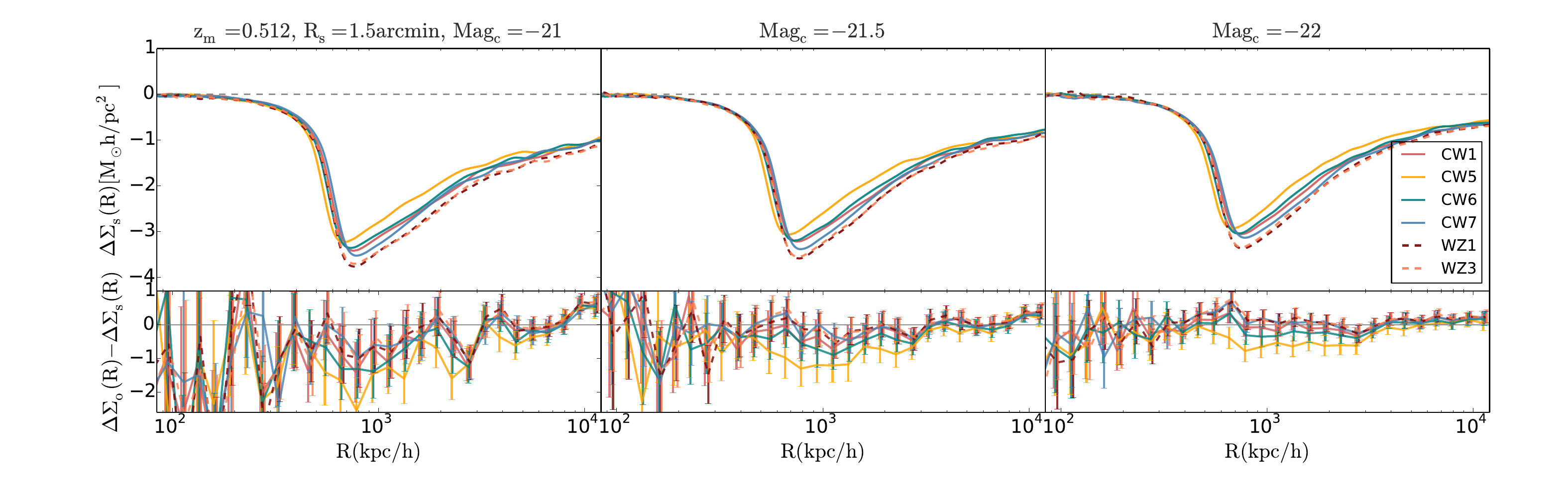}}
    \caption{Similar to fig.\ref{fig:6simu-1-new}, but with $\rm{R_s=1.5}$ arcmin.}\label{fig:6simu-1.5-new}
\end{figure*}

\begin{figure*}
    \subfigure{
    \includegraphics[width=0.93\textwidth, clip]{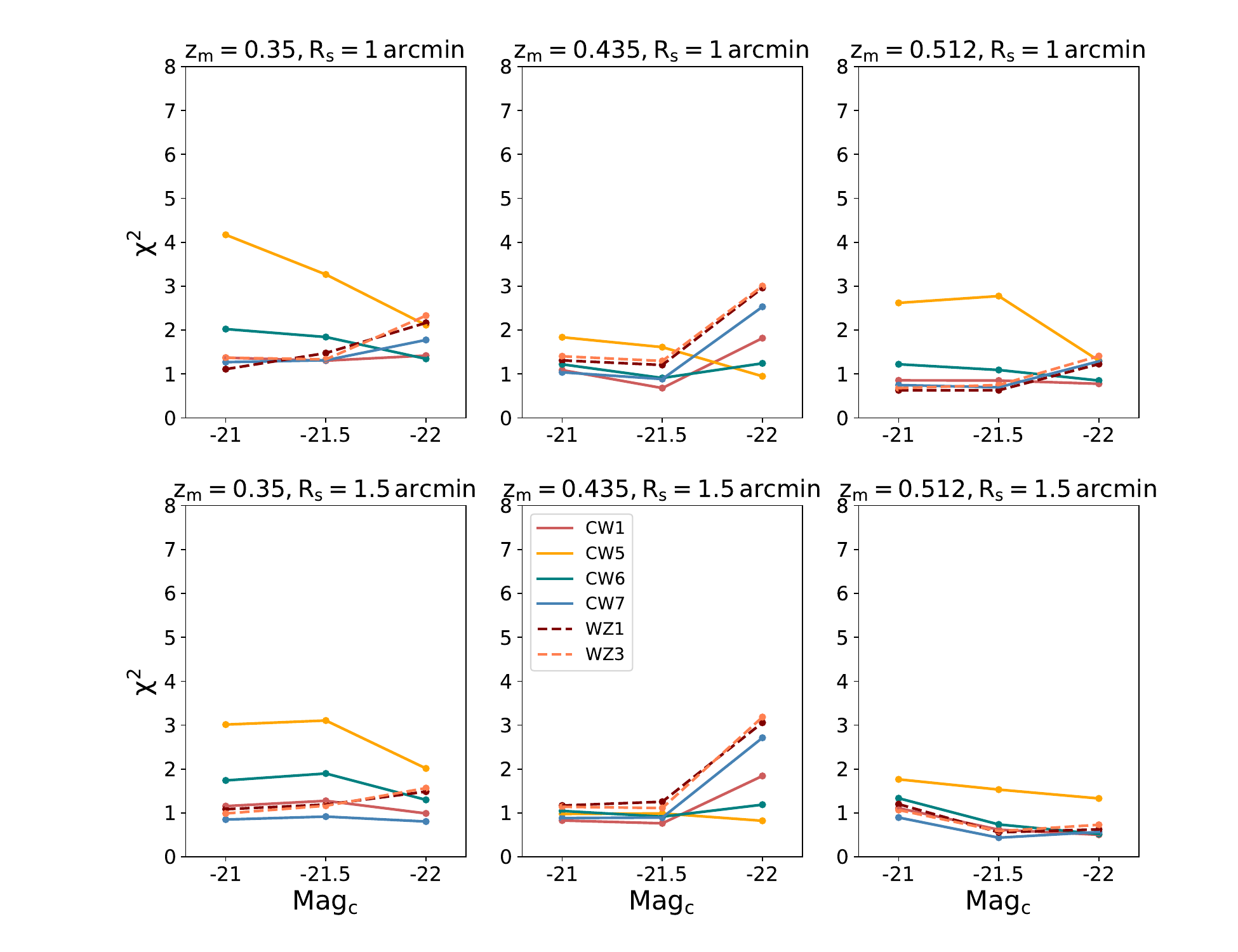}}
    \caption{Simalar to Fig.\ref{fig-kaf}, but with a new set of simulations described in \S\ref{new-simu}. CW5,6,7 all share the same $\rm{ \sigma_8}$ as that of CW1, and WZ3 has the same $\rm{\sigma_8}$ as that of WZ1. \label{fig-kaf-new}}
    \end{figure*}

All the procedures in \S\ref{section2} are  repeated for the four simulations.  The simulated lensing signals around the LDPs 
are compared with the observed signals in Fig.\ref{fig:6simu-1-new} and Fig.\ref{fig:6simu-1.5-new}, which are similar
to Fig.\ref{fig:6simu-1} and \ref{fig:6simu-1.5}. The lensing profiles $\rm{\Delta\Sigma_s(R)}$ for CW5,6,7 are  found to be close to
each other. Although smaller compared to Fig.\ref{fig:6simu-1} and \ref{fig:6simu-1.5}, discrepancies are still found for some
cosmologies between the simulated and observed signals in the lower panels.  Their corresponding $\rm{\chi^2}$ results are shown
in Fig.\ref{fig-kaf-new}. The CW5($\rm{w=-0.5}$) is found to have larger $\rm{\chi^2}$ than the others
in most cases.  The $\rm{\chi^2}$ of CW6($\rm{w=-0.8}$) seems to be slightly higher than the rest. These results are redshift-dependent, and the 
least distinguishable case is when $\rm{z_m=0.435}$ and $\rm{R_s=1.5\,arcmin}$, in which different models result in comparable $\rm{\chi^2}$.

\begin{table}[!htbp]
    \footnotesize
    \centering
     \caption{Simulation parameters.
    \label{Table2} }
    \begin{tabular}{c c c c c c c}
        \hline
        \hline
        Simulation  &$\mathrm{w_{de}}$      &$\sigma_8$  &$\Omega_c$  &$\Omega_b$   &h       &$\rm{n_s}$    \\
        \hline 
        CW5         &-0.5   &0.85          &0.223       &0.045        &0.71    &1\\
        CW6         &-0.8   &0.85          &0.223       &0.045        &0.71    &1\\
        CW7         &-1.2   &0.85          &0.223       &0.045        &0.71    &1\\
        \hline
        \hline
        Simulation  &$\mathrm{w_{de}}$      &$\mathrm{A_s}$       &$\Omega_c$  &$\Omega_b$   &h       &$\rm{n_s}$\\
        WZ3         &w(z)  &2.16e-9    &0.24188     &0.04525      &0.702   &0.966\\     
        \hline
    \end{tabular}
\end{table}

We note that our constraints on the dark energy equation of state is still preliminary, in the sense that we have fixed the values of the other cosmological parameters for simplicity. As a next step, we plan to vary the cosmological model with more parameters, and fix those that are best constrained by CMB. We also plan to measure again the LDP lensing signals using the Fourier\_Quad method \citep{2017ApJ...834....8Z,2018arXiv180802593Z}, which is significantly different from the Lensfit method used by the CFHTLenS team. Also, we are looking forward to giving detailed discussions on the redshift evolution of the LDP lensing signals with larger and deeper surveys\footnote{\url{https://www.darkenergysurvey.org},\url{https://www.desi.lbl.gov},\\ \url{https://www.lsst.org},\url{http://www.sdss3.org/surveys/boss.php.}}. \\

\acknowledgments{ACKNOWLEDGMENTS}\\

This work is based on observations obtained with MegaPrime/MegaCam, a
joint project of CFHT and CEA/DAPNIA, at the Canada-France-Hawaii
Telescope (CFHT) which is operated by the National Research Council
(NRC) of Canada, the Institut National des Sciences de l\'Univers of
the Centre National de la Recherche Scientifique (CNRS) of France, and
the University of Hawaii. This research used the facilities of the
Canadian Astronomy Data Centre operated by the National Research
Council of Canada with the support of the Canadian Space Agency.

We thank Gongbo Zhao for providing us parameter tables used for
setting the WZ set simulations, Yipeng Jing for providing us one of the CW1 simulation data, and Pengjie Zhang for useful comments. This work is supported by the National Key Basic Research and Development Program of China (No.2018YFA0404504), the National Key Basic Research Program of China (2015CB857001, 2015CB857002), the NSFC grants (11673016, 11433001, 11621303, 11773048, 11403071). JXH is supported by JSPS Grant-in-Aid for Scientific Research JP17K14271. JJZ is supported by China Postdoctoral Science Foundation 2018M632097.  LPF acknowledges the support from NSFC grants 11673018, 11722326 \& 11333001; STCSM grant 16ZR1424800 \& 188014066; and SHNU grant DYL201603.

\bibliography{ldp-lensing}

\begin{thebibliography}{}
\expandafter\ifx\csname natexlab\endcsname\relax\def\natexlab#1{#1}\fi

\bibitem[{{Barreira} {et~al.}(2017){Barreira}, {Bose}, {Li}, \&
  {Llinares}}]{2017JCAP...02..031B}
{Barreira}, A., {Bose}, S., {Li}, B., \& {Llinares}, C. 2017, \jcap, 2, 031

\bibitem[{{Behroozi} {et~al.}(2010){Behroozi}, {Conroy}, \&
  {Wechsler}}]{2010ApJ...717..379B}
{Behroozi}, P.~S., {Conroy}, C., \& {Wechsler}, R.~H. 2010, \apj, 717, 379

\bibitem[{{Berlind} \& {Weinberg}(2002)}]{2002ApJ...575..587B}
{Berlind}, A.~A., \& {Weinberg}, D.~H. 2002, \apj, 575, 587

\bibitem[{{Brouwer} {et~al.}(2018){Brouwer}, {Demchenko}, {Harnois-D{\'e}raps},
  {Bilicki}, {Heymans}, {Hoekstra}, {Kuijken}, {Alpaslan}, {Brough}, {Cai},
  {Costa-Duarte}, {Dvornik}, {Erben}, {Hildebrandt}, {Holwerda}, {Schneider},
  {Sif{\'o}n}, \& {van Uitert}}]{2018MNRAS.481.5189B}
{Brouwer}, M.~M., {Demchenko}, V., {Harnois-D{\'e}raps}, J., {et~al.} 2018,
  \mnras, 481, 5189

\bibitem[{{Ceccarelli} {et~al.}(2006){Ceccarelli}, {Padilla}, {Valotto}, \&
  {Lambas}}]{2006MNRAS.373.1440C}
{Ceccarelli}, L., {Padilla}, N.~D., {Valotto}, C., \& {Lambas}, D.~G. 2006,
  \mnras, 373, 1440

\bibitem[{{Chaves-Montero} {et~al.}(2016){Chaves-Montero}, {Angulo}, {Schaye},
  {Schaller}, {Crain}, {Furlong}, \& {Theuns}}]{2016MNRAS.460.3100C}
{Chaves-Montero}, J., {Angulo}, R.~E., {Schaye}, J., {et~al.} 2016, \mnras,
  460, 3100

\bibitem[{{Clampitt} \& {Jain}(2015)}]{2015MNRAS.454.3357C}
{Clampitt}, J., \& {Jain}, B. 2015, \mnras, 454, 3357

\bibitem[{{Colberg} {et~al.}(2005){Colberg}, {Sheth}, {Diaferio}, {Gao}, \&
  {Yoshida}}]{2005MNRAS.360..216C}
{Colberg}, J.~M., {Sheth}, R.~K., {Diaferio}, A., {Gao}, L., \& {Yoshida}, N.
  2005, \mnras, 360, 216

\bibitem[{{Conroy} {et~al.}(2009){Conroy}, {Gunn}, \&
  {White}}]{2009ApJ...699..486C}
{Conroy}, C., {Gunn}, J.~E., \& {White}, M. 2009, \apj, 699, 486

\bibitem[{{Conroy} {et~al.}(2006){Conroy}, {Wechsler}, \&
  {Kravtsov}}]{2006ApJ...647..201C}
{Conroy}, C., {Wechsler}, R.~H., \& {Kravtsov}, A.~V. 2006, \apj, 647, 201

\bibitem[{{Davies} {et~al.}(2018){Davies}, {Cautun}, \&
  {Li}}]{2018arXiv180308717D}
{Davies}, C.~T., {Cautun}, M., \& {Li}, B. 2018, ArXiv e-prints,
  arXiv:1803.08717

\bibitem[{{Elyiv} {et~al.}(2015){Elyiv}, {Marulli}, {Pollina}, {Baldi},
  {Branchini}, {Cimatti}, \& {Moscardini}}]{2015MNRAS.448..642E}
{Elyiv}, A., {Marulli}, F., {Pollina}, G., {et~al.} 2015, \mnras, 448, 642

\bibitem[{{Erben} {et~al.}(2013){Erben}, {Hildebrandt}, {Miller}, {van
  Waerbeke}, {Heymans}, {Hoekstra}, {Kitching}, {Mellier}, {Benjamin}, {Blake},
  {Bonnett}, {Cordes}, {Coupon}, {Fu}, {Gavazzi}, {Gillis}, {Grocutt}, {Gwyn},
  {Holhjem}, {Hudson}, {Kilbinger}, {Kuijken}, {Milkeraitis}, {Rowe},
  {Schrabback}, {Semboloni}, {Simon}, {Smit}, {Toader}, {Vafaei}, {van Uitert},
  \& {Velander}}]{2013MNRAS.433.2545E}
{Erben}, T., {Hildebrandt}, H., {Miller}, L., {et~al.} 2013, \mnras, 433, 2545

\bibitem[{{Friedrich} {et~al.}(2018){Friedrich}, {Gruen}, {DeRose}, {Kirk},
  {Krause}, {McClintock}, {Rykoff}, {Seitz}, {Wechsler}, {Bernstein}, {Blazek},
  {Chang}, {Hilbert}, {Jain}, {Kovacs}, {Lahav}, {Abdalla}, {Allam}, {Annis},
  {Bechtol}, {Benoit-L{\'e}vy}, {Bertin}, {Brooks}, {Carnero Rosell}, {Carrasco
  Kind}, {Carretero}, {Cunha}, {D'Andrea}, {da Costa}, {Davis}, {Desai},
  {Diehl}, {Dietrich}, {Drlica-Wagner}, {Eifler}, {Fosalba}, {Frieman},
  {Garc{\'{\i}}a-Bellido}, {Gaztanaga}, {Gerdes}, {Giannantonio}, {Gruendl},
  {Gschwend}, {Gutierrez}, {Honscheid}, {James}, {Jarvis}, {Kuehn},
  {Kuropatkin}, {Lima}, {March}, {Marshall}, {Melchior}, {Menanteau}, {Miquel},
  {Mohr}, {Nord}, {Plazas}, {Sanchez}, {Scarpine}, {Schindler}, {Schubnell},
  {Sevilla-Noarbe}, {Sheldon}, {Smith}, {Soares-Santos}, {Sobreira}, {Suchyta},
  {Swanson}, {Tarle}, {Thomas}, {Troxel}, {Vikram}, {Weller}, \& {DES
  Collaboration}}]{2017arXiv171005162F}
{Friedrich}, O., {Gruen}, D., {DeRose}, J., {et~al.} 2018, \prd, 98, 023508

\bibitem[{{Gruen} {et~al.}(2016){Gruen}, {Friedrich}, {Amara}, {Bacon},
  {Bonnett}, {Hartley}, {Jain}, {Jarvis}, {Kacprzak}, {Krause}, {Mana}, {Rozo},
  {Rykoff}, {Seitz}, {Sheldon}, {Troxel}, {Vikram}, {Abbott}, {Abdalla},
  {Allam}, {Armstrong}, {Banerji}, {Bauer}, {Becker}, {Benoit-L{\'e}vy},
  {Bernstein}, {Bernstein}, {Bertin}, {Bridle}, {Brooks}, {Buckley-Geer},
  {Burke}, {Capozzi}, {Carnero Rosell}, {Carrasco Kind}, {Carretero}, {Crocce},
  {Cunha}, {D'Andrea}, {da Costa}, {DePoy}, {Desai}, {Diehl}, {Dietrich},
  {Doel}, {Eifler}, {Neto}, {Fernandez}, {Flaugher}, {Fosalba}, {Frieman},
  {Gerdes}, {Gruendl}, {Gutierrez}, {Honscheid}, {James}, {Kuehn},
  {Kuropatkin}, {Lahav}, {Li}, {Lima}, {Maia}, {March}, {Martini}, {Melchior},
  {Miller}, {Miquel}, {Mohr}, {Nord}, {Ogando}, {Plazas}, {Reil}, {Romer},
  {Roodman}, {Sako}, {Sanchez}, {Scarpine}, {Schubnell}, {Sevilla-Noarbe},
  {Smith}, {Soares-Santos}, {Sobreira}, {Suchyta}, {Swanson}, {Tarle},
  {Thaler}, {Thomas}, {Walker}, {Wechsler}, {Weller}, {Zhang}, \&
  {Zuntz}}]{2016MNRAS.455.3367G}
{Gruen}, D., {Friedrich}, O., {Amara}, A., {et~al.} 2016, \mnras, 455, 3367

\bibitem[{{Gruen} {et~al.}(2018){Gruen}, {Friedrich}, {Krause}, {DeRose},
  {Cawthon}, {Davis}, {Elvin-Poole}, {Rykoff}, {Wechsler}, {Alarcon},
  {Bernstein}, {Blazek}, {Chang}, {Clampitt}, {Crocce}, {De Vicente}, {Gatti},
  {Gill}, {Hartley}, {Hilbert}, {Hoyle}, {Jain}, {Jarvis}, {Lahav}, {MacCrann},
  {McClintock}, {Prat}, {Rollins}, {Ross}, {Rozo}, {Samuroff}, {S{\'a}nchez},
  {Sheldon}, {Troxel}, {Zuntz}, {Abbott}, {Abdalla}, {Allam}, {Annis},
  {Bechtol}, {Benoit-L{\'e}vy}, {Bertin}, {Bridle}, {Brooks}, {Buckley-Geer},
  {Carnero Rosell}, {Carrasco Kind}, {Carretero}, {Cunha}, {D'Andrea}, {da
  Costa}, {Desai}, {Diehl}, {Dietrich}, {Doel}, {Drlica-Wagner}, {Fernandez},
  {Flaugher}, {Fosalba}, {Frieman}, {Garc{\'{\i}}a-Bellido}, {Gaztanaga},
  {Giannantonio}, {Gruendl}, {Gschwend}, {Gutierrez}, {Honscheid}, {James},
  {Jeltema}, {Kuehn}, {Kuropatkin}, {Lima}, {March}, {Marshall}, {Martini},
  {Melchior}, {Menanteau}, {Miquel}, {Mohr}, {Plazas}, {Roodman}, {Sanchez},
  {Scarpine}, {Schubnell}, {Sevilla-Noarbe}, {Smith}, {Smith}, {Soares-Santos},
  {Sobreira}, {Swanson}, {Tarle}, {Thomas}, {Vikram}, {Walker}, {Weller},
  {Zhang}, \& {DES Collaboration}}]{2017arXiv171005045G}
{Gruen}, D., {Friedrich}, O., {Krause}, E., {et~al.} 2018, \prd, 98, 023507

\bibitem[{{Guo} {et~al.}(2018){Guo}, {Yang}, \& {Lu}}]{2018ApJ...858...30G}
{Guo}, H., {Yang}, X., \& {Lu}, Y. 2018, \apj, 858, 30

\bibitem[{{Guo} {et~al.}(2016){Guo}, {Zheng}, {Behroozi}, {Zehavi}, {Chuang},
  {Comparat}, {Favole}, {Gottloeber}, {Klypin}, {Prada},
  {Rodr{\'{\i}}guez-Torres}, {Weinberg}, \& {Yepes}}]{2016MNRAS.459.3040G}
{Guo}, H., {Zheng}, Z., {Behroozi}, P.~S., {et~al.} 2016, \mnras, 459, 3040

\bibitem[{{Guo} \& {White}(2014)}]{2014MNRAS.437.3228G}
{Guo}, Q., \& {White}, S. 2014, \mnras, 437, 3228

\bibitem[{{Guo} {et~al.}(2010){Guo}, {White}, {Li}, \&
  {Boylan-Kolchin}}]{2010MNRAS.404.1111G}
{Guo}, Q., {White}, S., {Li}, C., \& {Boylan-Kolchin}, M. 2010, \mnras, 404,
  1111

\bibitem[{{Han} {et~al.}(2012){Han}, {Frenk}, {Eke}, {Gao}, {White},
  {Boyarsky}, {Malyshev}, \& {Ruchayskiy}}]{2012MNRAS.427.1651H}
{Han}, J., {Frenk}, C.~S., {Eke}, V.~R., {et~al.} 2012, \mnras, 427, 1651

\bibitem[{{Hartlap} {et~al.}(2007){Hartlap}, {Simon}, \&
  {Schneider}}]{2007A&A...464..399H}
{Hartlap}, J., {Simon}, P., \& {Schneider}, P. 2007, \aap, 464, 399

\bibitem[{{Hearin} {et~al.}(2013){Hearin}, {Zentner}, {Berlind}, \&
  {Newman}}]{2013MNRAS.433..659H}
{Hearin}, A.~P., {Zentner}, A.~R., {Berlind}, A.~A., \& {Newman}, J.~A. 2013,
  \mnras, 433, 659

\bibitem[{{Heymans} {et~al.}(2012){Heymans}, {Van Waerbeke}, {Miller}, {Erben},
  {Hildebrandt}, {Hoekstra}, {Kitching}, {Mellier}, {Simon}, {Bonnett},
  {Coupon}, {Fu}, {Harnois D{\'e}raps}, {Hudson}, {Kilbinger}, {Kuijken},
  {Rowe}, {Schrabback}, {Semboloni}, {van Uitert}, {Vafaei}, \&
  {Velander}}]{2012MNRAS.427..146H}
{Heymans}, C., {Van Waerbeke}, L., {Miller}, L., {et~al.} 2012, \mnras, 427,
  146

\bibitem[{{Hildebrandt} {et~al.}(2012){Hildebrandt}, {Erben}, {Kuijken}, {van
  Waerbeke}, {Heymans}, {Coupon}, {Benjamin}, {Bonnett}, {Fu}, {Hoekstra},
  {Kitching}, {Mellier}, {Miller}, {Velander}, {Hudson}, {Rowe}, {Schrabback},
  {Semboloni}, \& {Ben{\'{\i}}tez}}]{2012MNRAS.421.2355H}
{Hildebrandt}, H., {Erben}, T., {Kuijken}, K., {et~al.} 2012, \mnras, 421, 2355

\bibitem[{{Hoyle} \& {Vogeley}(2002)}]{2002ApJ...566..641H}
{Hoyle}, F., \& {Vogeley}, M.~S. 2002, \apj, 566, 641

\bibitem[{{Huterer} \& {Turner}(1999)}]{1999PhRvD..60h1301H}
{Huterer}, D., \& {Turner}, M.~S. 1999, \prd, 60, 081301

\bibitem[{{Jennings} {et~al.}(2013){Jennings}, {Li}, \&
  {Hu}}]{2013MNRAS.434.2167J}
{Jennings}, E., {Li}, Y., \& {Hu}, W. 2013, \mnras, 434, 2167

\bibitem[{{Jing} {et~al.}(1998){Jing}, {Mo}, \&
  {B{\"o}rner}}]{1998ApJ...494....1J}
{Jing}, Y.~P., {Mo}, H.~J., \& {B{\"o}rner}, G. 1998, \apj, 494, 1

\bibitem[{{Jing} {et~al.}(2007){Jing}, {Suto}, \& {Mo}}]{2007ApJ...657..664J}
{Jing}, Y.~P., {Suto}, Y., \& {Mo}, H.~J. 2007, \apj, 657, 664

\bibitem[{{Komatsu} {et~al.}(2014){Komatsu}, {Bennett}, {Barnes}, {Bean},
  {Bennett}, {Dor{\'e}}, {Dunkley}, {Gold}, {Greason}, {Halpern}, {Hill},
  {Hinshaw}, {Jarosik}, {Kogut}, {Komatsu}, {Larson}, {Limon}, {Meyer},
  {Nolta}, {Odegard}, {Page}, {Peiris}, {Smith}, {Spergel}, {Tucker}, {Verde},
  {Weiland}, {Wollack}, \& {Wright}}]{2014PTEP.2014fB102K}
{Komatsu}, E., {Bennett}, C.~L., {Barnes}, C., {et~al.} 2014, Progress of
  Theoretical and Experimental Physics, 2014, 06B102

\bibitem[{{Lavaux} \& {Wandelt}(2012)}]{2012ApJ...754..109L}
{Lavaux}, G., \& {Wandelt}, B.~D. 2012, \apj, 754, 109

\bibitem[{{Leauthaud} {et~al.}(2011){Leauthaud}, {Tinker}, {Behroozi}, {Busha},
  \& {Wechsler}}]{2011ApJ...738...45L}
{Leauthaud}, A., {Tinker}, J., {Behroozi}, P.~S., {Busha}, M.~T., \&
  {Wechsler}, R.~H. 2011, \apj, 738, 45

\bibitem[{Lewis {et~al.}(2000)Lewis, Challinor, \& Lasenby}]{Lewis:1999bs}
Lewis, A., Challinor, A., \& Lasenby, A. 2000, Astrophys. J., 538, 473

\bibitem[{{Li}(2011)}]{2011MNRAS.411.2615L}
{Li}, B. 2011, \mnras, 411, 2615

\bibitem[{{Miller} {et~al.}(2013){Miller}, {Heymans}, {Kitching}, {van
  Waerbeke}, {Erben}, {Hildebrandt}, {Hoekstra}, {Mellier}, {Rowe}, {Coupon},
  {Dietrich}, {Fu}, {Harnois-D{\'e}raps}, {Hudson}, {Kilbinger}, {Kuijken},
  {Schrabback}, {Semboloni}, {Vafaei}, \& {Velander}}]{2013MNRAS.429.2858M}
{Miller}, L., {Heymans}, C., {Kitching}, T.~D., {et~al.} 2013, \mnras, 429,
  2858

\bibitem[{{Nadathur} {et~al.}(2015){Nadathur}, {Hotchkiss}, {Diego}, {Iliev},
  {Gottl{\"o}ber}, {Watson}, \& {Yepes}}]{2015MNRAS.449.3997N}
{Nadathur}, S., {Hotchkiss}, S., {Diego}, J.~M., {et~al.} 2015, \mnras, 449,
  3997

\bibitem[{{Neyrinck}(2008)}]{2008MNRAS.386.2101N}
{Neyrinck}, M.~C. 2008, \mnras, 386, 2101

\bibitem[{{Niikura} {et~al.}(2015){Niikura}, {Takada}, {Okabe}, {Martino}, \&
  {Takahashi}}]{2015PASJ...67..103N}
{Niikura}, H., {Takada}, M., {Okabe}, N., {Martino}, R., \& {Takahashi}, R.
  2015, \pasj, 67, 103

\bibitem[{{Padilla} {et~al.}(2005){Padilla}, {Ceccarelli}, \&
  {Lambas}}]{2005MNRAS.363..977P}
{Padilla}, N.~D., {Ceccarelli}, L., \& {Lambas}, D.~G. 2005, \mnras, 363, 977

\bibitem[{{Peacock}(1999)}]{1998...Peacock}
{Peacock}, J.~A. 1999, {Cosmological Physics}, 704

\bibitem[{{Perlmutter} {et~al.}(1999){Perlmutter}, {Aldering}, {Goldhaber},
  {Knop}, {Nugent}, {Castro}, {Deustua}, {Fabbro}, {Goobar}, {Groom}, {Hook},
  {Kim}, {Kim}, {Lee}, {Nunes}, {Pain}, {Pennypacker}, {Quimby}, {Lidman},
  {Ellis}, {Irwin}, {McMahon}, {Ruiz-Lapuente}, {Walton}, {Schaefer}, {Boyle},
  {Filippenko}, {Matheson}, {Fruchter}, {Panagia}, {Newberg}, {Couch}, \&
  {Project}}]{1999ApJ...517..565P}
{Perlmutter}, S., {Aldering}, G., {Goldhaber}, G., {et~al.} 1999, \apj, 517,
  565

\bibitem[{{Planck Collaboration} {et~al.}(2014){Planck Collaboration}, {Ade},
  {Aghanim}, {Alves}, {Armitage-Caplan}, {Arnaud}, {Ashdown},
  {Atrio-Barandela}, {Aumont}, {Aussel}, \& et~al.}]{2014A&amp;A...571A...1P}
{Planck Collaboration}, {Ade}, P.~A.~R., {Aghanim}, N., {et~al.} 2014, \aap,
  571, A1

\bibitem[{{Platen} {et~al.}(2007){Platen}, {van de Weygaert}, \&
  {Jones}}]{2007MNRAS.380..551P}
{Platen}, E., {van de Weygaert}, R., \& {Jones}, B.~J.~T. 2007, \mnras, 380,
  551

\bibitem[{{Ramos} {et~al.}(2011){Ramos}, {Pellegrini}, {Benoist}, {da Costa},
  {Maia}, {Makler}, {Ogando}, {de Simoni}, \& {Mesquita}}]{2011AJ....142...41R}
{Ramos}, B.~H.~F., {Pellegrini}, P.~S., {Benoist}, C., {et~al.} 2011, \aj, 142,
  41

\bibitem[{{Riess} {et~al.}(1998){Riess}, {Filippenko}, {Challis},
  {Clocchiatti}, {Diercks}, {Garnavich}, {Gilliland}, {Hogan}, {Jha},
  {Kirshner}, {Leibundgut}, {Phillips}, {Reiss}, {Schmidt}, {Schommer},
  {Smith}, {Spyromilio}, {Stubbs}, {Suntzeff}, \&
  {Tonry}}]{1998AJ....116.1009R}
{Riess}, A.~G., {Filippenko}, A.~V., {Challis}, P., {et~al.} 1998, \aj, 116,
  1009

\bibitem[{{Rodr{\'{\i}}guez-Puebla} {et~al.}(2015){Rodr{\'{\i}}guez-Puebla},
  {Avila-Reese}, {Yang}, {Foucaud}, {Drory}, \& {Jing}}]{2015ApJ...799..130R}
{Rodr{\'{\i}}guez-Puebla}, A., {Avila-Reese}, V., {Yang}, X., {et~al.} 2015,
  \apj, 799, 130

\bibitem[{{Rodr{\'{\i}}guez-Puebla} {et~al.}(2017){Rodr{\'{\i}}guez-Puebla},
  {Primack}, {Avila-Reese}, \& {Faber}}]{2017MNRAS.470..651R}
{Rodr{\'{\i}}guez-Puebla}, A., {Primack}, J.~R., {Avila-Reese}, V., \& {Faber},
  S.~M. 2017, \mnras, 470, 651

\bibitem[{{S{\'a}nchez} {et~al.}(2017){S{\'a}nchez}, {Clampitt}, {Kovacs},
  {Jain}, {Garc{\'{\i}}a-Bellido}, {Nadathur}, {Gruen}, {Hamaus}, {Huterer},
  {Vielzeuf}, {Amara}, {Bonnett}, {DeRose}, {Hartley}, {Jarvis}, {Lahav},
  {Miquel}, {Rozo}, {Rykoff}, {Sheldon}, {Wechsler}, {Zuntz}, {Abbott},
  {Abdalla}, {Annis}, {Benoit-L{\'e}vy}, {Bernstein}, {Bernstein}, {Bertin},
  {Brooks}, {Buckley-Geer}, {Carnero Rosell}, {Carrasco Kind}, {Carretero},
  {Crocce}, {Cunha}, {D'Andrea}, {da Costa}, {Desai}, {Diehl}, {Dietrich},
  {Doel}, {Evrard}, {Fausti Neto}, {Flaugher}, {Fosalba}, {Frieman},
  {Gaztanaga}, {Gruendl}, {Gutierrez}, {Honscheid}, {James}, {Krause}, {Kuehn},
  {Lima}, {Maia}, {Marshall}, {Melchior}, {Plazas}, {Reil}, {Romer}, {Sanchez},
  {Schubnell}, {Sevilla-Noarbe}, {Smith}, {Soares-Santos}, {Sobreira},
  {Suchyta}, {Tarle}, {Thomas}, {Walker}, {Weller}, \& {DES
  Collaboration}}]{2017MNRAS.465..746S}
{S{\'a}nchez}, C., {Clampitt}, J., {Kovacs}, A., {et~al.} 2017, \mnras, 465,
  746

\bibitem[{{Simha} {et~al.}(2012){Simha}, {Weinberg}, {Dav{\'e}}, {Fardal},
  {Katz}, \& {Oppenheimer}}]{2012MNRAS.423.3458S}
{Simha}, V., {Weinberg}, D.~H., {Dav{\'e}}, R., {et~al.} 2012, \mnras, 423,
  3458

\bibitem[{{Springel}(2005)}]{2005MNRAS.364.1105S}
{Springel}, V. 2005, \mnras, 364, 1105

\bibitem[{{Springel} \& {Hernquist}(2002)}]{2002MNRAS.333..649S}
{Springel}, V., \& {Hernquist}, L. 2002, \mnras, 333, 649

\bibitem[{{Taylor} {et~al.}(2013){Taylor}, {Joachimi}, \&
  {Kitching}}]{2013MNRAS.432.1928T}
{Taylor}, A., {Joachimi}, B., \& {Kitching}, T. 2013, \mnras, 432, 1928

\bibitem[{{Vale} \& {Ostriker}(2004)}]{2004MNRAS.353..189V}
{Vale}, A., \& {Ostriker}, J.~P. 2004, \mnras, 353, 189

\bibitem[{{Vale} \& {Ostriker}(2006)}]{2006MNRAS.371.1173V}
---. 2006, \mnras, 371, 1173

\bibitem[{{van den Bosch} {et~al.}(2007){van den Bosch}, {Yang}, {Mo},
  {Weinmann}, {Macci{\`o}}, {More}, {Cacciato}, {Skibba}, \&
  {Kang}}]{2007MNRAS.376..841V}
{van den Bosch}, F.~C., {Yang}, X., {Mo}, H.~J., {et~al.} 2007, \mnras, 376,
  841

\bibitem[{{Wechsler} \& {Tinker}(2018)}]{2018arXiv180403097W}
{Wechsler}, R.~H., \& {Tinker}, J.~L. 2018, ArXiv e-prints, arXiv:1804.03097

\bibitem[{{Weinberg} {et~al.}(2013){Weinberg}, {Mortonson}, {Eisenstein},
  {Hirata}, {Riess}, \& {Rozo}}]{2013PhR...530...87W}
{Weinberg}, D.~H., {Mortonson}, M.~J., {Eisenstein}, D.~J., {et~al.} 2013,
  \physrep, 530, 87

\bibitem[{{Yang} {et~al.}(2003){Yang}, {Mo}, \& {van den
  Bosch}}]{2003MNRAS.339.1057Y}
{Yang}, X., {Mo}, H.~J., \& {van den Bosch}, F.~C. 2003, \mnras, 339, 1057

\bibitem[{{Yang} {et~al.}(2012){Yang}, {Mo}, {van den Bosch}, {Zhang}, \&
  {Han}}]{2012ApJ...752...41Y}
{Yang}, X., {Mo}, H.~J., {van den Bosch}, F.~C., {Zhang}, Y., \& {Han}, J.
  2012, \apj, 752, 41

\bibitem[{{Yang} {et~al.}(2018){Yang}, {Zhang}, {Wang}, {Liu}, {Lu}, {Li},
  {Shi}, {Jing}, {Mo}, {van den Bosch}, {Kang}, {Cui}, {Guo}, {Li}, {Lim},
  {Lu}, {Luo}, {Wei}, \& {Yang}}]{2018ApJ...860...30Y}
{Yang}, X., {Zhang}, Y., {Wang}, H., {et~al.} 2018, \apj, 860, 30

\bibitem[{{Zehavi} {et~al.}(2011){Zehavi}, {Zheng}, {Weinberg}, {Blanton},
  {Bahcall}, {Berlind}, {Brinkmann}, {Frieman}, {Gunn}, {Lupton}, {Nichol},
  {Percival}, {Schneider}, {Skibba}, {Strauss}, {Tegmark}, \&
  {York}}]{2011ApJ...736...59Z}
{Zehavi}, I., {Zheng}, Z., {Weinberg}, D.~H., {et~al.} 2011, \apj, 736, 59

\bibitem[{{Zhang} {et~al.}(2017){Zhang}, {Zhang}, \&
  {Luo}}]{2017ApJ...834....8Z}
{Zhang}, J., {Zhang}, P., \& {Luo}, W. 2017, \apj, 834, 8

\bibitem[{{Zhang} {et~al.}(2018){Zhang}, {Dong}, {Li}, {Li}, {Li}, {Liu},
  {Luo}, {Fu}, {Li}, \& {Fan}}]{2018arXiv180802593Z}
{Zhang}, J., {Dong}, F., {Li}, H., {et~al.} 2018, ArXiv e-prints,
  arXiv:1808.02593

\bibitem[{{Zhao} {et~al.}(2016){Zhao}, {Tao}, {Liang}, {Kitaura}, \&
  {Chuang}}]{2016MNRAS.459.2670Z}
{Zhao}, C., {Tao}, C., {Liang}, Y., {Kitaura}, F.-S., \& {Chuang}, C.-H. 2016,
  \mnras, 459, 2670

\bibitem[{{Zhao} {et~al.}(2017){Zhao}, {Raveri}, {Pogosian}, {Wang},
  {Crittenden}, {Handley}, {Percival}, {Beutler}, {Brinkmann}, {Chuang},
  {Cuesta}, {Eisenstein}, {Kitaura}, {Koyama}, {L'Huillier}, {Nichol}, {Pieri},
  {Rodriguez-Torres}, {Ross}, {Rossi}, {S{\'a}nchez}, {Shafieloo}, {Tinker},
  {Tojeiro}, {Vazquez}, \& {Zhang}}]{2017NatAs...1..627Z}
{Zhao}, G.-B., {Raveri}, M., {Pogosian}, L., {et~al.} 2017, Nature Astronomy,
  1, 627

\bibitem[{{Zheng} {et~al.}(2005){Zheng}, {Berlind}, {Weinberg}, {Benson},
  {Baugh}, {Cole}, {Dav{\'e}}, {Frenk}, {Katz}, \&
  {Lacey}}]{2005ApJ...633..791Z}
{Zheng}, Z., {Berlind}, A.~A., {Weinberg}, D.~H., {et~al.} 2005, \apj, 633, 791

\bibitem[{{Zu} \& {Mandelbaum}(2016)}]{2016MNRAS.457.4360Z}
{Zu}, Y., \& {Mandelbaum}, R. 2016, \mnras, 457, 4360

\end{thebibliography}

\end{document}